\documentclass[12pt,aps,prd,preprint,tightenlines,superscriptaddress,showpacs,nofootinbib]{revtex4-1}
\usepackage{graphicx}

\usepackage{amsmath} 
\usepackage{graphicx}
   \usepackage{multirow}

\usepackage{epstopdf}
\newcommand{\PRE}[1]{{#1}} 

\newcommand{\be}{\begin{equation}}
\newcommand{\ee}{\end{equation}}
\newcommand{\bea}{\begin{eqnarray}}
\newcommand{\eea}{\end{eqnarray}}

\def\tev{\, {\rm TeV}}
\def\gev{\, {\rm GeV}}
\def\mev{\, {\rm MeV}}
\def\kev{\, {\rm keV}}

\newcommand{\gsim}{\lower.7ex\hbox{$\;\stackrel{\textstyle>}{\sim}\;$}}
\newcommand{\lsim}{\lower.7ex\hbox{$\;\stackrel{\textstyle<}{\sim}\;$}}

\newcommand{\kg}{{\rm kg}}
\newcommand{\da}{{\rm day}}

\usepackage{enumerate}
\usepackage{amssymb,amsmath}

\begin{document}

\preprint{UH-511-1278-2017}
\preprint{CETUP2016-012}

\title{
\PRE{\vspace*{1.3in}}
\textsc{A Study of Dark Matter and QCD-Charged Mediators in the Quasi-Degenerate
Regime}
\PRE{\vspace*{0.1in}}
}

\author{Andrew Davidson}
\affiliation{\mbox{Department of Physics, University of North Florida, Jacksonville, FL  32224, USA}
}
\affiliation{\mbox{Department of Physics and Astronomy, University of Utah, Salt Lake City, UT  84112, USA}
}

\author{Chris Kelso}
\affiliation{\mbox{Department of Physics, University of North Florida, Jacksonville, FL  32224, USA}
}

\author{Jason Kumar}
\affiliation{\mbox{Department of Physics and Astronomy, University of
Hawai'i, Honolulu, HI 96822, USA }
}

\author{Pearl Sandick}
\affiliation{\mbox{Department of Physics and Astronomy, University of Utah, Salt Lake City, UT  84112, USA}
}

\author{Patrick Stengel \PRE{\vspace*{.1in}} }
\affiliation{\mbox{Department of Physics and Astronomy, University of
Hawai'i, Honolulu, HI 96822, USA }
}
\affiliation{Michigan Center for Theoretical Physics, Department of Physics, University of Michigan, Ann Arbor, MI 48109, USA,
\PRE{\vspace*{.1in}}
}

\begin{abstract}

We study a scenario in which the only light new particles are a Majorana fermion
dark matter candidate and one or more QCD-charged scalars, which couple to light quarks.
This scenario  has several interesting phenomenological features if the new particles
are nearly degenerate in mass.   In particular, LHC
searches for the light scalars have reduced sensitivity, since the visible and
invisible products tend to be softer.  Moreover, dark matter-scalar co-annihilation can
allow even relatively heavy dark matter candidates to be consistent thermal relics.
Finally, the dark matter nucleon scattering cross section is enhanced in the
quasi-degenerate limit, allowing direct detection experiments to use both spin-independent
and spin-dependent scattering to probe regions of parameter space beyond those
probed by the LHC.  Although this scenario has broad application,
we phrase this study in terms of the MSSM, in the limit where the only light sparticles
are a bino-like dark matter candidate and light-flavored squarks.
\end{abstract}

\maketitle


\section{Introduction}

A well-studied framework for dark matter (DM) interactions with the Standard Model (SM) is the scenario
of a SM-singlet Majorana fermion, which interacts with SM fermions via the exchange of SM-charged scalar mediators~\cite{Chang:2013oia,An:2013xka,Bai:2013iqa,Papucci:2014iwa,Garny:2014waa}.
The canonical example of this scenario arises in the MSSM in the case where the lightest supersymmetric
particle (LSP) is a bino, which interacts with SM fermions through the exchange of sfermions.  But there exist
other examples of this scenario, including, for example, WIMPless dark matter models~\cite{Feng:2008ya,Feng:2008mu,Barger:2010ng,Fukushima:2011df}.
In this work, we study the phenomenology of this scenario in the limit
where DM couples to light quarks through the exchange of a QCD-charged scalar whose mass is nearly degenerate with the
DM particle.  We will find that this scenario has unique features that manifest themselves in direct detection,
in dark matter co-annihilation in the early Universe, and in searches at the LHC.

It is well-known that if the LSP is mostly bino-like, then there are obstacles to its viability as a thermal relic,
as well as to its discovery at direct detection experiments.  In particular, if flavor violation is minimal, then
$s$-wave dark matter annihilation is chirality-suppressed, and cannot deplete the relic density enough to
remain consistent with cosmological observations.
Velocity-dependent contributions to the annihilation cross section are suppressed by a factor of $\sim 10$ at the time of
freeze-out.
The spin-independent (SI) dark matter-nucleon scattering
cross section is also chirality-suppressed in the non-relativistic limit, reducing the sensitivity of direct detection experiments to these models.
Both the annihilation and scattering cross sections are further suppressed by the heavy sfermion masses necessary for consistency with LHC searches.
Indeed, the
``bulk" region of the CMSSM parameter space with an acceptable thermal relic density has long been excluded (see, e.g.~\cite{Ellis:2012nv}).
The suppression of the dark matter annihilation and scattering
cross sections can be alleviated if flavor violation is non-minimal, as is the case if there is non-trivial sfermion mixing.
This ``Incredible Bulk" region of
parameter space has been studied in~\cite{Fukushima:2014yia,Kelso:2014qja}.
Again, these results generalize beyond the implementation of this scenario in the MSSM.

As pointed out in~\cite{Kelso:2014qja}, there is an interesting region of parameter space in which the dark matter and a charged
mediator are nearly degenerate.
In portions of this region of parameter space, the relic density is depleted by co-annihilation of dark matter with
the charged mediator in the early Universe, yielding a thermal relic density that is consistent with observation.
This scenario has also been well-studied in the context of the CMSSM, where the charged mediators are typically $\tilde \tau$ or $\tilde t$~\cite{Ellis:1998kh, Ellis:1999mm, Boehm:1999bj, Ellis:2001nx}, but there has been less study of the case where the
charged mediator couples to $u$-, $d$-, or $s$-quarks.
Additionally, in this limit, the tight collider bounds
on the mass of the QCD-charged mediator
can be relaxed.
Finally, the dark matter-nucleon scattering cross sections,
both spin-independent (SI) and spin-dependent (SD),
are significantly enhanced when the dark matter and a charged mediator are nearly degenerate.
As a result, direct detection experiments can potentially probe regions of parameter space for
which the charged mediators are well beyond the reach of the LHC, even if dark matter-nucleon
scattering is largely spin- or velocity-dependent. In addition, higher dimensional effective operators, which are
usually suppressed by the mass splitting between the dark matter and the mediator, can provide the
dominant contribution to dark matter-nucleon scattering in cases where the sfermion mixing is small.

The outline of this paper is as follows.  In section II, we introduce the model and describe
general features of dark matter-nucleon scattering, LHC constraints, and dark matter annihilation/co-annihilation.
In section III, we apply this analysis to some benchmark examples and determine regions of parameter
space for which the dark matter can be a thermal relic, and regions to which direct detection experiments
are sensitive.  We conclude with a discussion of our results in section IV.

\section{The Model}

We consider a model in which the dark matter particle is a SM gauge singlet Majorana fermion $\chi$ which
interacts with a SM quark $q$ through the exchange of scalars $\tilde q_{L,R}$ which
are charged under both $SU(3)_{qcd}$ and $U(1)_{em}$.  The DM-SM interaction Lagrangian
can be written as
\bea
{\cal L} &=& \sum _{q=u,d,s,c} \lambda_{Lq} ( \bar \chi P_L q ) \tilde{q}_L^* + \lambda_{Rq} ( \bar \chi P_R q ) \tilde{q}_R^* + h.c. \, ,
\label{eq:Interaction}
\eea
where we focus only on the first two generations.
We assume that $\chi$ is absolutely stable, because it is the lightest particle charged under some discrete unbroken $Z_2$ symmetry.
Since SM particles are assumed to be neutral under this new discrete symmetry, the $\tilde q_{L,R}$ are thus also necessarily odd
under $Z_2$.  Thus, $\chi$ and $\tilde q_{L,R}$ have the quantum numbers of the MSSM bino and left-/right-squarks, respectively.
We may then consider this Lagrangian as a simplified model for the MSSM scenario in which the only light sparticles are a bino-like
LSP and some number of light squarks.  In this case we have $\lambda_{Lq,Rq} = \sqrt{2} g' Y_{L,R}$, where $g'$ is the hypercharge
coupling and $Y_{L,R}$ are the hypercharges of the left- and right-handed quarks, respectively.  But this model has wider applicability,
and, more generally, the couplings $\lambda_{Lq,Rq}$ can assume any values, constrained only by perturbativity.

In general, the chiral eigenstates $\tilde q_{L,R}$ can mix, and can be expressed in terms of
the mass eigenstates $\tilde q_{1,2}$ as
\bea
\tilde{q}_L &=&  \tilde{q}_1 \cos \alpha + \tilde{q}_2 \sin \alpha ,
\nonumber\\
\tilde{q}_R &=&  - \tilde{q}_1 \sin \alpha + \tilde{q}_2 \cos \alpha ,
\eea
where the possible $CP$-violating phase has been absorbed into the relative phase of the
couplings $\lambda_{Lq,Rq}$.
We denote the mass eigenvalues of the $\tilde q_{1,2}$ states by $m_{\tilde q_{1,2}}$, and assume
without loss of generality that $m_\chi < m_{\tilde q_1} \leq m_{\tilde q_2}$.

If the dark matter couples to several quarks, then mixing between scalars that couple to different
quarks can contribute to flavor-changing neutral currents (FCNCs).  There are tight experimental constraints on
such FCNCs, so for simplicity, we assume that any mixing between the $\tilde q_L$ and $\tilde q_R$ does not mix
generations.  Note that although each scalar pair can have a different mixing angle, for simplicity of notation and
presentation, we assume that the mixing angles are all identical.  This simplifying assumption
will not affect our results qualitatively.

Because the scalars $\tilde q_{L,R}$ are charged under SM gauge groups, gauge invariance requires that they couple
to SM gauge bosons.  Although these interactions will not be relevant for dark matter-nucleon scattering or
annihilation,  they can affect the dark matter relic density
through co-annihilation processes.  These Lagrangian terms are given in Appendix~\ref{app:scalInt}.

The interactions in Eq.~\ref{eq:Interaction} will also lead to corrections to the SM fermion-photon vertex
through one-loop diagrams in which $\chi$ and $\tilde q_{1,2}$ run in the loop~\cite{Fukushima:2013efa,Fukushima:2014yia,Kumar:2016gxq}.
These corrections can be constrained by
collider experiments, and by precision measurements of fermion magnetic and electric dipole moments.  Although these
constraints can be very tight if the fermion is a lepton, current data does not rule out any interesting regions of parameter space
in the case where the fermion is a quark~\cite{Fukushima:2013efa,Fukushima:2014yia,Kumar:2016gxq}.

\subsection{Dark Matter-Nucleon Scattering}

DM-nucleon scattering is mediated by $s$- and $u$-channel exchange of scalar mediators.  In the non-relativistic
limit, the DM-quark scattering matrix element can be derived from a linear combination of effective
contact operators defined at the weak scale,
\bea
{\cal O}_q &=& \sum_{i=1}^7 {\cal O}_{qi},
\eea
where the dimension-6 contact operators are given by~\cite{Fukushima:2011df}
\bea
{\cal O}_{q1} &=& \alpha_{q1} (\bar \chi \gamma^\mu \gamma^5 \chi) (\bar q \gamma_\mu q) ,
\nonumber\\
{\cal O}_{q2} &=& \alpha_{q2} (\bar \chi \gamma^\mu \gamma^5 \chi) (\bar q \gamma_\mu \gamma^5 q) ,
\nonumber\\
{\cal O}_{q3} &=& \alpha_{q3} (\bar \chi  \chi) (\bar q  q) ,
\nonumber\\
{\cal O}_{q4} &=& \alpha_{q4} (\bar \chi \gamma^5 \chi) (\bar q  \gamma^5 q) ,
\nonumber\\
{\cal O}_{q5} &=& \alpha_{q5} (\bar \chi \chi) (\bar q  \gamma^5 q) ,
\nonumber\\
{\cal O}_{q6} &=& \alpha_{q6} (\bar \chi \gamma^5 \chi) (\bar q   q) ,
\eea
with
\bea
\alpha_{q1} &= &
-\left[{|\lambda_L^2| \over 8}
\left( {\cos^2 \alpha \over m_{\tilde q_1}^2 -m_\chi^2}
+ {\sin^2 \alpha \over m_{\tilde q_2}^2 -m_\chi^2} \right)
-{|\lambda_R^2| \over 8}
\left({\cos^2 \alpha \over m_{\tilde q_2}^2 -m_\chi^2}
+ {\sin^2 \alpha \over m_{\tilde q_1}^2 -m_\chi^2} \right) \right] ,
\nonumber\\
\alpha_{q2} &= &
\left[{|\lambda_L^2| \over 8}
\left( {\cos^2 \alpha \over m_{\tilde q_1}^2 -m_\chi^2}
+ {\sin^2 \alpha \over m_{\tilde q_2}^2 -m_\chi^2} \right)
+{|\lambda_R^2| \over 8}
\left({\cos^2 \alpha \over m_{\tilde q_2}^2 -m_\chi^2}
+ {\sin^2 \alpha \over m_{\tilde q_1}^2 -m_\chi^2} \right) \right] ,
\nonumber\\
\alpha_{q3,4} &=& {Re(\lambda_L \lambda_R^*) \over 4}(\cos \alpha \sin \alpha)
\left[{1\over m_{\tilde q_1}^2 -m_\chi^2} -{1\over m_{\tilde q_2}^2 -m_\chi^2} \right] ,
\nonumber\\
\alpha_{q5,6} &=& {\imath Im(\lambda_L \lambda_R^*) \over 4}(\cos \alpha \sin \alpha)
\left[{1\over m_{\tilde q_1}^2 -m_\chi^2} -{1\over m_{\tilde q_2}^2 -m_\chi^2} \right] .
\label{eq:Coefficients}
\eea

Higher dimension contact operators arise from expanding the scalar propagators in powers of
the quark and dark matter momenta.  In general, these operators are subdominant because their
contributions to the dark matter-nucleon scattering matrix element are suppressed by additional factors of
$m_N m_\chi / (m_{{\tilde q}_1}^2 - m_\chi^2) \sim (m_N/ 2 \Delta m)$.
But the most important of these is the dimension-8 twist-2 operator~\cite{Drees:1993bu}
\bea
{\cal O}_{q7} &=& \alpha_{q7 } (\imath \bar \chi \gamma^\mu \partial^\nu \chi )
\left[ \left(\frac{\imath}{2} \right)
\left(\bar q \gamma_\mu \partial_\nu q + \bar q \gamma_\nu \partial_\mu q - \frac{1}{2} g_{\mu \nu} \bar q \gamma_\alpha \gamma^\alpha q  \right) \right] ,
\eea
where
\bea
\alpha_{q7} &=&
\frac{| \lambda_L^2 |}{4} \left[ { \cos^2 \alpha \over (m_{\tilde{q}_1}^2 - m_\chi^2)^2 }
+ { \sin^2 \alpha \over (m_{\tilde{q}_2}^2 - m_\chi^2)^2 } \right]
+ \frac{| \lambda_R^2 |}{4}
\left[ { \cos^2 \alpha \over (m_{\tilde{q}_2}^2 - m_\chi^2)^2 }
+ { \sin^2 \alpha \over (m_{\tilde{q}_1}^2 - m_\chi^2)^2 } \right] .
\eea
Because ${\cal O}_{q7}$ can mediate velocity-independent SI scattering even in the chiral limit, it
can provide an important contribution in the limit of small mixing ($\alpha \rightarrow 0$), especially
if $\Delta m / m_\chi \ll 1$.

Note that all of the effective
operator coefficients are enhanced in the limit $m_\chi / m_{\tilde q_1} \rightarrow 1$; in this limit,
the propagator of the mediator goes nearly on-shell.
However, if
$m_\chi$ and $m_{\tilde q_1}$ are sufficiently degenerate, then the expansion in contact operators no longer
provides a good approximation, because the dependence of the scalar propagators on the momentum transfer is no longer
small.  But since $|\overrightarrow{q}| \lesssim {\cal O}(100~\mev)$ for all relevant target nuclei,
the contact operator expansion will be valid provided $m_{\tilde q_1} - m_\chi \gtrsim 1~\gev$.

In Eq.~\ref{eq:Coefficients} we have neglected the decay width in the scalar propagators.
This is justified because the decay width for the scalar $\tilde q$ is necessarily
small compared to the mass splitting, $\Delta m = m_{\tilde q} - m_\chi$, due to the final state phase space
suppression.  In the nearly-degenerate limit the denominator of the scalar propagator goes as
$m_\chi^2 - m_{\tilde q}^2 +\imath \Gamma m_{\tilde q} \sim -m_\chi (2\Delta m -\imath \Gamma)$, implying
that the decay width term can indeed be neglected.

The operators
${\cal O}_{q2}$ and ${\cal O}_{q3,7}$ yield velocity-independent terms in the scattering matrix
element which are spin-dependent and spin-independent, respectively.  The remaining operators
generate only velocity-suppressed terms in the scattering matrix element.
The terms in the matrix element contributed by ${\cal O}_{q4}$ lead to spin-dependent scattering (without a
coherent scattering enhancement) and are suppressed by a factor $v^2$ (other velocity-suppressed operators
contribute matrix element terms suppressed only by $v$), so we can essentially ignore this operator.  In contrast, ${\cal O}_{q1}$ also leads to velocity-suppressed scattering, but it can still be significant as discussed below.
The coefficients $\alpha_{q5,6}$ vanish if the DM-SM interaction is
$CP$-invariant, and the coefficients $\alpha_{q3-6}$ are suppressed for the case of minimal flavor
violation (MFV).  As one might expect, the coefficients $\alpha_{q3-6}$ are also suppressed if
$m_{\tilde q_1} \simeq m_{\tilde q_2}$; these coefficients are only non-vanishing if there is non-trivial
scalar mixing, but if the scalar mediators are degenerate then the mixing angle can be rotated away by
a change of the mass eigenstate basis.

For simplicity, we assume that DM-SM interactions are $CP$-invariant, and $\lambda_{qL} / \lambda_{qR} \sim
{\cal O}(1)$.  In this case, since ${\cal O}_{q4-6}$ are essentially irrelevant, only four of the
effective operators are important for direct detection:
\begin{itemize}
\item{${\cal O}_{q1}$: this operator provides a velocity-suppressed contribution to the scattering
matrix element, which can be important if $\alpha$ is small.
}
\item{${\cal O}_{q2}$: this operator provides the dominant contribution to the SD scattering
matrix element.}
\item{${\cal O}_{q3}$: this operator provides the dominant contribution to the SI scattering
matrix element, unless $\alpha$ is small.
}
\item{${\cal O}_{q7}$: this operator provides the dominant contribution to the SI scattering
matrix element if $\alpha$ is small.
}
\end{itemize}
Note that there is no interference between the leading contributions of the first two effective
operators with any others~\cite{Kumar:2013iva}.  But operators ${\cal O}_{q3}$ and ${\cal O}_{q7}$ necessarily
interfere~\cite{Anand:2013yka}.

Since the coefficients $\alpha_{qi}$ are defined at the weak scale (which we may take as $\sim m_Z$),
one must determine the coefficients $\alpha_{qi}' (\mu)$
which arise from the RG evolution of the
effective contact operators from the weak scale down to a lower scale $\mu$.
We can determine the running of the coefficients from a high scale $\mu_H$ to a low scale $\mu_L$ using the results in~\cite{Hill:2014yxa}, yielding
\bea
\alpha_{q1}' (\mu_L) &=& \alpha_{q1}' (\mu_H) ,
\nonumber\\
 \left(
   \begin{array}{c}
     \alpha_{u2}' (\mu_L) \\
     \alpha_{d2}' (\mu_L) \\
     \alpha_{s2}' (\mu_L) \\
   \end{array}
 \right)
 &=& U
 \left(
                     \begin{array}{ccc}
                       A & 0 & 0 \\
                       0 & 1 & 0 \\
                       0 & 0 & 1 \\
                     \end{array}
                   \right)
 U^{-1}
 \left(
   \begin{array}{c}
     \alpha_{u2}' (\mu_H) \\
     \alpha_{d2}' (\mu_H) \\
     \alpha_{s2}' (\mu_H) \\
   \end{array}
 \right)
 ,
\nonumber\\
\alpha_{q3}' (\mu_L) &=& {m_q (\mu_L) \over m_q (\mu_H)} \alpha_{q3}' (\mu_H) ,
\nonumber\\
\alpha_{q7}' (\mu_L) &=& r(0)  \alpha_{q7}' (\mu_H)
+\sum_{q'=light} \frac{1}{n_f} \left[\frac{16 r(n_f) +3n_f}{16 +3n_f} -r(0) \right]\alpha_{q'7}' (\mu_H) ,
\eea
where
\bea
U &=& \left(
        \begin{array}{ccc}
          {1 \over \sqrt{3}} & {1 \over \sqrt{3}} & {1 \over \sqrt{3}} \\
          {1 \over \sqrt{6}} & {1 \over \sqrt{6}} & -\sqrt{2 \over 3} \\
          {1 \over \sqrt{2}} & -{1 \over \sqrt{2}} & 0 \\
        \end{array}
      \right) ,
      \nonumber\\
A &=& \exp \left[ {2n_f \over \pi \beta_0} \left( \alpha_s (\mu_H) - \alpha_s (\mu_L) \right) + {\cal O}(\alpha_s^2 ) \right] ,
\nonumber\\
r(t) &=& \left(\frac{\alpha_s(\mu_L)}{\alpha_s(\mu_H)} \right)^{-\frac{1}{2\beta_0}\left(\frac{64}{9} + \frac{4}{3}t \right)}.
\eea
The strong coupling constant $\alpha_s$ and the quark mass parameter $m_q$ are evaluated at the
appropriate scale in the $\overline{MS}$ scheme~\cite{Hill:2014yxa}, and  $\beta_0 = 11 - (2/3)n_f $
where $n_f$ is the number of relevant quark flavors.  Using the boundary condition $\alpha_{qi}' (m_Z) \equiv \alpha_{qi}$,
the coefficients may then be run straightforwardly to the nucleon scale (which we may take as $\mu \sim 1-2~\gev$).  Note
that the running of the operators ${\cal O}_{q2,7}$ changes slightly as one crosses the $b$-quark threshold.
The operator ${\cal O}_{q1}$ evolves trivially below the weak scale because the quark vector
current is protected by gauge invariance.  The flavor non-singlet axial vector quark current is also
scale-independent,  but the flavor singlet axial vector
current has a weak dependence on scale.  RG evolution has a much larger effect on the operator
${\cal O}_{q3}$.

The velocity-independent contributions to the scattering matrix element
generated by operators ${\cal O}_{q2}$ and ${\cal O}_{q3,q7}$
can be classified as spin-dependent and spin-independent, respectively.
For these operators, the differential DM-nucleus scattering cross sections are then given by:
\bea \label{eqn:DDXsec}
{d\sigma_A^{{\cal O}(SD)} \over dE_R } &=& {16 \mu_A^2 \over \pi E_R^{max}  } \left({J+1 \over J} \right)
\left(\sum_q \alpha_{q2}' \left( \langle S_p \rangle \Delta_q^{(p)} + \langle S_n \rangle \Delta_q^{(n)} \right)  \right)^2
|F_{{\cal O}_2} (E_R)|^2 ,
\nonumber\\
{d\sigma_A^{{\cal O}(SI)} \over dE_R } &=& {4 \mu_A^2 \over \pi E_R^{max}  }
\left(\sum_q \alpha_{q3}' \left( Z B_q^{p(S)} + (A-Z) B_q^{n(S)} \right)
\right.
\nonumber\\
&\,& \left.
+\frac{3}{4} m_N m_\chi \sum_q \alpha_{q7}' \left( Z B_q^{p(T2)} + (A-Z) B_q^{n(T2)} \right)
\right)^2 |F_{{\cal O}_3} (E_R)|^2 ,
\eea
where
$\mu_A = m_A m_\chi /(m_A + m_\chi)$ is the DM-nucleus reduced mass and $m_A$ is the mass of the target
nucleus. $E_R^{max} = 2\mu_A^2 v^2 / m_A$ is the maximum nuclear recoil energy which is kinematically allowed
if $v$ is the relative velocity.  The $F_{{\cal O}_i} (E_R)$ are nuclear form factors (which we obtain from~\cite{Anand:2013yka}),
and the $B_q$ and $\Delta_q$ are nucleon form factors.

Operator ${\cal O}_{q1}$ couples dark matter to vector quark currents.
However, the nuclear response cannot be expressed simply in
terms of SI and/or SD nuclear form factors.
For example, there is an additional term that
arises from coupling to the orbital angular momentum of the nucleons.
The complete expression for the DM-nucleus scattering cross section can be found in~\cite{Anand:2013yka},
and we use that expression, and the associated nuclear response functions, in our subsequent
numerical calculations.

The nucleon form factors for a vector quark current interaction are completely determined by
gauge invariance.
The other nucleon form factors have some uncertainty, especially for the scalar interaction.
For the scalar nucleon form factor, we will make a conservative estimate regarding the strangeness content of the nucleon,
and adopt the following values as a benchmark~\cite{Kelso:2014qja}:
\bea
B_u^{p(S)} &=& B_d^{n(S)} =9.85 ,
\nonumber\\
B_d^{p(S)} &=& B_u^{n(S)} =6.77 ,
\nonumber\\
B_{s}^{p,n(S)} &=& 0.499 .
\eea
The effect on direct detection sensitivity of varying the strangeness content of the nucleon is further discussed
in~\cite{Kelso:2014qja}.

For the twist-2 operator, we will for simplicity use nucleon form factors given in~\cite{Hill:2014yxa}:
\bea
B_u^{p(T2)} &=& B_d^{n(T2)} =0.40 ,
\nonumber\\
B_u^{n(T2)} &=& B_d^{p(T2)} = 0.22 ,
\nonumber\\
B_s^{p,n(T2)} &=&  0.02 .
\eea

For the axial-vector spin nucleon form factors, we will for simplicity use the values used in~\cite{Dienes:2013xya}:
\bea
\Delta_u^{(p)} &=& \Delta_d^{(n)} = 0.787 ,
\nonumber\\
\Delta_u^{(n)} &=& \Delta_d^{(p)} = -0.319 ,
\nonumber\\
\Delta_s^{(p,n)} &=&  -0.040 .
\eea
We note that, given these nucleon form factors, the DM-nucleon scattering rates for models with either mass degenerate $u$- and $d$-type squarks or mass degenerate $u$-, $d$- and $s$-type squarks will be nearly identical. While the relative smallness of the form factors for strange quarks is manifest for each effective operator we consider, we again note that the scalar nucleon form factor for strange quarks can be considerably larger than the value we use. With a larger strangeness content in the nucleon, we would expect an ${\cal O}(1)$ enhancement to the scattering cross section contribution arising from the scalar effective operator.

\subsection{LHC Constraints on the Mediator Mass}

This scenario can be probed at the LHC, utilizing searches for the production of the mediator
through QCD processes.  A variety of such searches have been performed in the context of
the MSSM, in regions of parameter space where the only light strongly-coupled superparticles
are squarks.  But the constraints on the mediator mass derived from these SUSY searches
can be generalized to other models that realize this scenario.

As discussed in~\cite{Kelso:2014qja}, LHC searches for squark pair production in the scenario of eight degenerate light-flavor squarks
and decoupled gluinos generically exclude squark masses $m_{\tilde{q}} \lsim { \cal O } ( 1.4 \tev)$ for $m_{{\chi}} \sim {\cal O }
( 100 \gev )$~\cite{Sirunyan:2017cwe}.
In scenarios with one non-degenerate light-flavor squark which is significantly heavier than the neutralino LSP (assuming all other sparticles
are decoupled), mass constraints weaken considerably to $m_{\tilde{q}_1} \gsim { \cal O } (1.0 \tev)$~\cite{Sirunyan:2017cwe}.

If, alternatively, one light-flavor squark is nearly degenerate in mass with the LSP, the low transverse momenta of the squark decay products and the low missing transverse energy of the final state make extracting the squark pair production signal from the QCD background difficult, almost independent
of the mass scale. In order to probe a more compressed spectrum with $ m_{\tilde{q}_1} - m_{{\chi}} \lsim 25 \gev $, event selection can include the presence of initial-state radiation (ISR) jets, which can be used to identify signal events and will boost the transverse missing energy of the final state~\cite{ATLAS:2017dnw,Sirunyan:2017kiw}. Although
recent analysis does not study the specific interpretation relevant for the benchmarks studied in this work, we note that, for production of eight degenerate light-flavor squarks, $m_{\tilde{q}} \lsim 700 \gev$ is excluded~\cite{ATLAS:2017dnw}. Also, assuming a spectrum with a nearly degenerate sbottom and LSP, the mass exclusion weakens to $m_{\tilde{b}_1} \lsim 600 \gev$~\cite{Sirunyan:2017kiw}. While reinterpretation of these results for our simplified model is beyond the scope of this work, we consider any scenario with $m_{\tilde{q}} \lsim 400 \gev$
to be ruled out by LHC.  But a dedicated analysis of current data could improve this bound by ${\cal O}(200) \gev$, assuming the LHC sensitivity to the production of a single light-flavor squark is similar to that of the sbottom search.

We have focused on the scenario in which the only new accessible particles are the dark matter and the scalar mediators.
But, for example, the production of first generation squarks in SUSY models with a light gluino ($m_{\tilde g} \sim \cal{O} ( \tev)$) will be
enhanced though $t$-channel gluino production~\cite{Mahbubani:2012qq}, resulting in an increased sensitivity to such models at the LHC.
Thus, LHC constraints may be more severe for specific models in which there are additional light QCD-coupled new particles, beyond those assumed
in the simplified model that we consider.

\subsection{Dark Matter Annihilation and Co-Annihilation}

The cross section for the dark matter annihilation process $\chi \chi \rightarrow \bar q q$ in this model has been
computed in~\cite{Fukushima:2014yia}, in the limit $m_q / m_\chi \ll 1$.  As expected, the $s$-wave contribution
to the annihilation matrix element vanishes in the chiral limit as $\alpha \rightarrow 0$.  This follows from the
fact that an $s$-wave initial state of two identical fermions must have $J=0$, implying that the $\bar q q$ final state
must contain a fermion and anti-fermion of the same helicity; such a final state can only arise from an interaction that
mixes left-handed and right-handed Weyl spinors.  Thus, if $\alpha \sim {\cal O}(1)$, the $s$-wave annihilation cross section may be
substantial both at the time of thermal freeze-out and in the current epoch.  But if $\alpha \ll 1$, then dark matter
annihilation at freeze-out may be dominated by $p$-wave annihilation, which is suppressed by a factor $v^2 \sim 0.1$;
in the current epoch one finds $v^2 \sim 10^{-6}$, so $p$-wave annihilation today would be negligible.

If $m_\chi / m_{\tilde q_1} \sim 1$, then both $\tilde q_1$ and $\chi$ will be abundant in the early Universe at the
time of dark matter thermal freeze-out.  Because either one of these light supersymmetric particles can convert into the other via
scattering with relativistic SM particles, one can determine the dark matter relic density by computing the evolution of the
total density of both species, including the effects of DM and scalar annihilation as well as DM-scalar co-annihilation.
Of these processes, only DM annihilation is chirality-suppressed; in the $\alpha \rightarrow 0$ limit, annihilation and
co-annihilation processes involving the scalars can thus  play an important role
in depleting the thermal relic density in the early Universe.
But in the present epoch, when $\tilde q_1$ is no longer abundant, the annihilation/co-annihilation processes involving
the scalars are negligible.

We can identify three classes of processes which are included in the total dark matter annihilation rate with relative contributions approximately determined by $\Delta m = m_{\tilde q_1} - m_\chi $ in different regions of $(m_\chi, \Delta m)$ parameter  space:
\begin{itemize}
\item The non-degenerate region, where $\Delta m$ is large enough that one can  ignore co-annihilation contributions.
The process $ \chi \chi \rightarrow \bar q q$ dominates the depletion of the relic density.
\item The nearly degenerate region, where $\Delta m$ is small enough that processes like $ \chi \tilde q \to g q$
are significant, but large enough that processes such as $\tilde q^* \tilde q \rightarrow gg$ are insignificant, as a result of the
Boltzmann-suppression of the abundance of the heavier state.
\item The degenerate region, where $\Delta m$ is so small that the Boltzmann-suppression of the heavier state in
      the early Universe is negligible.  In this limit, channels such as
      $\tilde q^* \tilde q \to g g$ and $\tilde q^* \tilde q \to g Z$ yield the dominant annihilation contributions.
\end{itemize}
As we shall see, for squark masses and mass splittings allowed by LHC constraints, the correct relic density can only be reproduced in the degenerate region of the $(m_\chi, \Delta m)$ parameter space. Since the relevant squark-squark annihilation processes with purely electroweak final states will be suppressed by the light-flavor quark masses or electroweak gauge couplings,
and processes with a QCD final state will be enhanced due to the strong gauge couplings, the dominant contribution to the depletion of the relic density comes from $\tilde q^* \tilde q \to g g$, with a cross section given by~\cite{deSimone:2014pda}
\bea
\langle \sigma v ( \tilde q^* \tilde q \to g g ) \rangle = {7 g_s^4 N_{\tilde q} \over 432 \pi m_{\tilde q}^2 } \left[ N_{\tilde q} + {\exp \left( \Delta m / T \right) \over 3 \left( 1 + \Delta m / m_\chi \right)^{3/2}}  \right]^{-2},
\eea
after summing over $N_{\tilde q}$ light mass-degenerate squarks and noting the temperature near freezeout is typically $T \sim m_\chi /25$.
Because this is a purely QCD process, the cross section remains the same independent of squark flavor or $L$-$R$ mixing angle.
Note that for $\tilde q^* \tilde q \to g g$, and in general, reproducing the correct relic density requires
$m_{\chi}$ or $m_{\tilde q}$ to be light enough that the relevant annihilation cross sections
are not suppressed by the mass scale.

 Also, in the degenerate region of parameter space, the cross sections for all processes will decrease with the introduction of additional light squarks due to the dilution of the total number density across individual species, as demonstrated by the $N_{\tilde q}$ dependence in the $\tilde q^* \tilde q \to g g$ cross section.
This dilution effect becomes less pronounced as $\Delta m $ increases, but in general the overall annihilation rate will decrease with the addition of mass degenerate light species unless the new annihilation channels associated with the additional fields are efficient.
Thus, it is worth noting that the relic density can actually increase as additional
scalars are made light.

In addition to their couplings to dark matter, the scalar mediators necessarily couple to the $\gamma$, $g$, $Z$ and $W^{\pm}$,
as described in Appendix~\ref{app:scalInt}.
But beyond this minimal set of interactions, one can also write renormalizable gauge-invariant interactions of the scalars with each other and
with the SM Higgs.  Such terms will arise generically within the MSSM Lagrangian. In particular, there are $D$-term contributions to the squark-squark and squark-Higgs interactions, which, unlike the contributions arising from the superpotential or the soft SUSY-breaking trilinear terms, are not proportional to the light-flavor quark masses.  We find that the inclusion of $D$-term squark-squark and squark-Higgs interactions does not significantly alter
the relic density calculation, so our results for the simplified model we discuss  are also valid within the framework of the MSSM.

Note that although annihilation/co-annihilation processes involving light scalars in the initial state may be relevant to the depletion of the dark matter
relic density in the early Universe, they are not relevant to indirect detection in the current epoch.
However, for the purposes of indirect detection, in addition to the $s$-wave
annihilation process $\chi \chi \rightarrow \bar q q$, one should also consider the internal bremsstrahlung
process ($\chi \chi \rightarrow \bar q q \gamma$)~\cite{Bringmann:2007nk},
and the process where dark matter annihilates to monoenergetic photons through a
one-loop diagram ($\chi \chi \rightarrow \gamma \gamma,\gamma Z$)~\cite{Bergstrom:1997fh,Bern:1997ng,Ullio:1997ke}.
The internal bremsstrahlung process is particularly useful for indirect detection when the dark matter
and charged scalar are nearly degenerate, because the photon spectrum becomes very hard (almost line-like) due to
a collinear divergence.
The importance of these processes for the case in which dark matter couples to leptons was recently considered in~\cite{Kumar:2016cum},
but those results generalize to the case where dark matter couples to quarks.

\section{Results}

In this section, we specialize to the case where the dark matter candidate is the bino of the MSSM,
which couples to light SM quarks through squark exchange.  We thus set
$\lambda_{L,R} = \sqrt{2} g' Y_{L,R}$, where $g'$ is the hypercharge coupling constant and
$Y_{L,R}$ are the left- and right-handed quark hypercharges.
The DM-nucleus scattering cross sections and DM thermal
relic density will thus depend only on $\alpha$, $m_\chi$, and on the masses of the light squarks, $m_{\tilde q_i}$.
The DM relic abundance is calculated with MicrOMEGAs version 4.3.4~\cite{Belanger:2001fz, Belanger:2004yn, Belanger:2014vza},
while the DM-nucleus scattering cross sections are calculated using the formalism of~\cite{Anand:2013yka}.

In contrast to previous studies, we explore the co-annihilation parameter space for light-flavor squarks and allow for $L$-$R$ squark mixing.  Specifically,  we will focus on five benchmark scenarios:
\begin{itemize}
\item{Benchmark A) a single light squark, $\tilde u_1$;}
\item{Benchmark B) a single light squark, $\tilde s_1$;}
\item{Benchmark C) two light degenerate squarks, $\tilde u_1$ and $\tilde d_1$;}
\item{Benchmark D) two light degenerate squarks, $\tilde u_1$ and $\tilde u_2$;}
\item{Benchmark E) three light degenerate squarks, $\tilde u_1$, $\tilde d_1$ and $\tilde s_1$.}
\end{itemize}
For Benchmarks C, D, and E, we assume that the light squarks are degenerate, so, for all five benchmarks we may denote the light squark mass as $m_{\tilde q}$ and define $\Delta m \equiv m_{\tilde q} - m_\chi >0$.
For any particular benchmark, the relative strength of the coupling of dark matter to each light quark
is determined; henceforth, for simplicity of presentation, we will refer the effective operators relevant
for direct detection as ${\cal O}_i$, and dispense with the $q$ subscript.

Of course, gauge-invariance under $SU(2)_L$ implies that one cannot keep only one squark light while
absolutely decoupling all other squarks.  Moreover precision electroweak constraints on the $\rho$ parameter
imply that, absent canceling corrections from other new physics, the mass splitting between any of the squarks
within the same generation cannot be too large (see, for example,\cite{Drees:1990dx}).
But for practical
purposes, this will not affect our results, since we are largely focused on the regime in which the dark matter
and the lightest squark are nearly degenerate.  Provided the mass splitting between squarks is significantly larger
than the mass splitting between the dark matter and the lightest squark, dark matter-nucleon scattering will be
dominated by exchange of the lightest squark, and co-annihilation processes in the early Universe will be affected
dominantly by the lightest squark.

In fact, corrections to the $\rho$ parameter arising from a scalar loop roughly scale as
\bea
\delta \rho &\sim& \frac{c^2}{16\pi^2 } {\cal O}\left(\frac{\delta m^2}{m_{Z}^2} \right) ,
\label{eqn:rhomix}
\eea
where $\delta m$ is the mass splitting between two squarks, $c$ is the coupling between the scalars 
and a weak gauge boson, and we take
$\delta  m \ll m_{{\tilde q}_1}$.  Corrections to $\rho$ will be at the percent level, consistent with experimental
constraints, provided $\delta m \lesssim {\cal O}(100~\gev)$.  For the region of parameter space
of greatest interest, we will find $\Delta m \equiv m_{{\tilde q}_1} - m_\chi \ll 100~\gev$, implying that
indeed we are justified in ignoring the presence of the heavier squarks for the purpose of direct detection
and co-annihilation in the early Universe.

\subsection{Direct Detection Prospects}
\label{sec:ddresults}

We now consider the sensitivity of direct detection experiments to this class of models, in
which the DM and charged scalar mediator(s) have a small mass splitting.  The DM-nucleus scattering cross sections are nearly identical for Benchmarks C and E because the
nucleon form factors that we have used for $s$-quark interactions are relatively small; $s$-quark
interactions will only be significant if the coupling to $u$ and $d$ quarks are suppressed.
Note, however, that the
presence of an additional light $s$-squark can have a significant effect on the dark matter thermal
relic density, as will be discussed in Section~\ref{sec:rdresults}.

\begin{figure}[ht]
\includegraphics*[width=0.49 \textwidth]{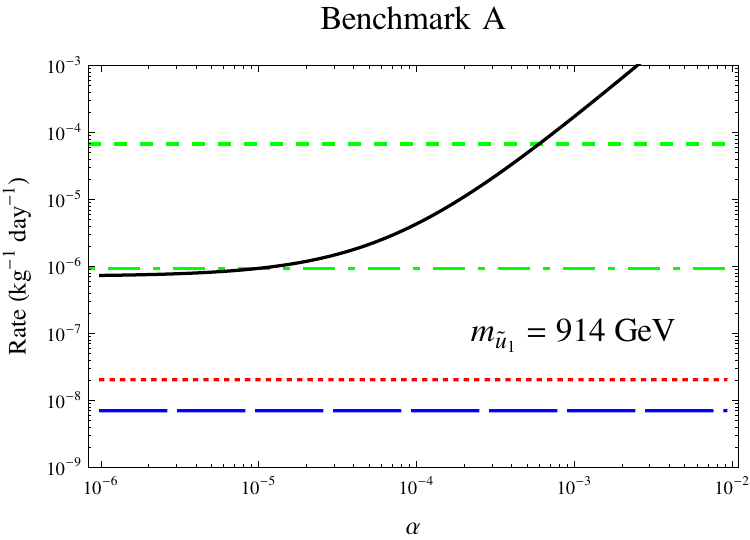}
\includegraphics*[width=0.49 \textwidth]{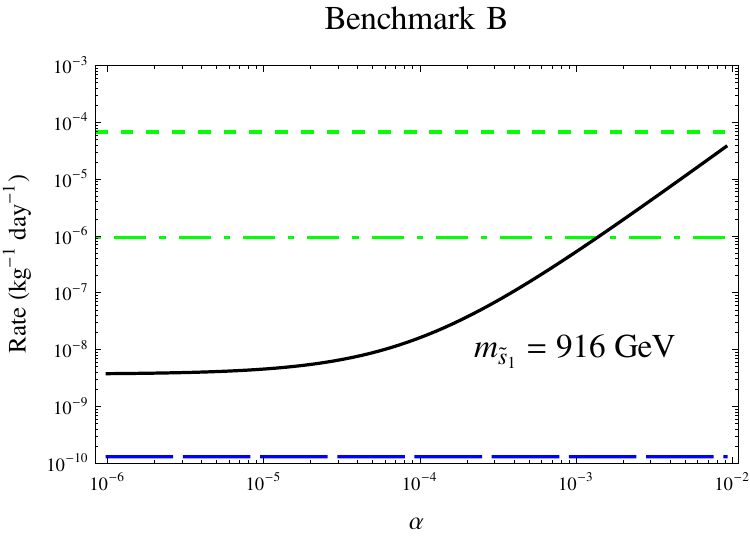} \\ \vspace{2mm}
\includegraphics*[width=0.49 \textwidth]{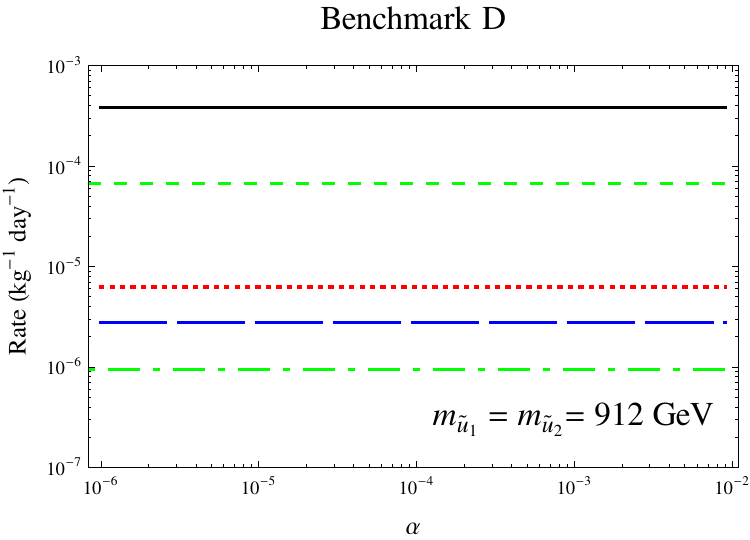}
\includegraphics*[width=0.49 \textwidth]{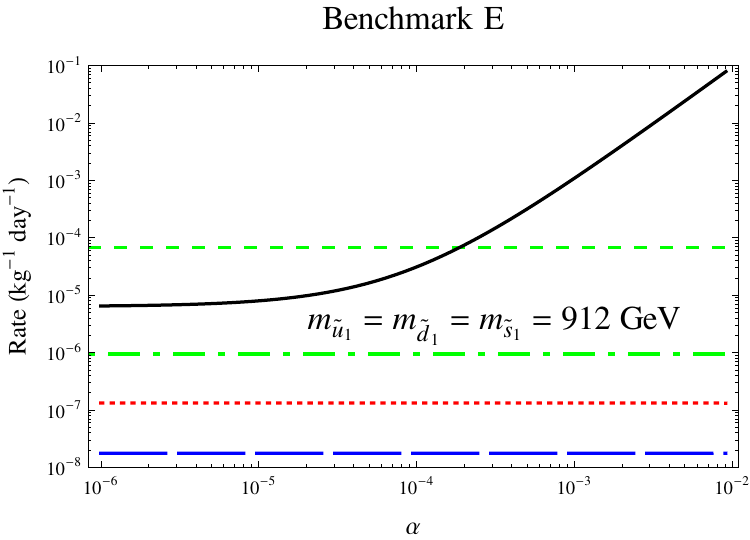}
\caption{Event rate at a xenon-based detector (assuming the recoil energy range of XENON1T)
as a function of $\alpha$ for operators ${\cal O}_1$ (dotted red), ${\cal O}_2$ (long-dashed blue) and ${\cal O}_{3,7}$ (solid black)
for a fixed choice of $m_\chi = 900 \gev $.  Shown are Benchmark A (upper left panel),
Benchmark B (upper right panel), Benchmark D (lower left panel)
and Benchmark E (lower right panel).  All other squarks have a negligible effect.
The short-dashed and dash-dotted green lines indicate the current 90\% CL sensitivity of
XENON1T~\cite{Aprile:2017iyp} and the estimated 90\% CL sensitivity of
LZ~\cite{Mount:2017qzi}, respectively.  For Benchmarks A, B and E,  operator ${\cal O}_7$ begins to 
dominate the SI scattering rate for $\alpha \lesssim 10^{-4-5}$, where the 
solid black line is flat.
}
\label{fig:ratecomp}
\end{figure}

In Figure~\ref{fig:ratecomp}, we show the rate of scattering events at a xenon-based detector
(in the energy range $5\kev - 40\kev$ used by XENON1T) resulting from
interactions mediated by operators ${\cal O}_{1,2,3,7}$, as a function of $\alpha$, with
$m_{\chi} = 900~\gev$.   We show Benchmarks A, B, D, and E, and in each case the masses of the light
squarks are chosen so that the thermal relic density will be consistent with observations (within the range
$912-916 \gev$).  We omit Benchmark C because, as previously mentioned, the scattering rates are very similar to Benchmark E.
At leading level, the sets of operators $\{ {\cal O}_{1} \}$, $\{ {\cal O}_{2} \}$, and $\{ {\cal O}_{3}, {\cal O}_{7} \}$, do not interfere with each other,
so we plot the events rates arising from each set of operators separately.
The short-dashed and dash-dotted green lines indicate the current 90\% CL sensitivity of XENON1T ($7\times 10^{-5}~\kg^{-1} \da^{-1}$\cite{Aprile:2017iyp}) and
the future estimated 90\% CL sensitivity of LZ ($9\times 10^{-7}~\kg^{-1} \da^{-1}$~\cite{Mount:2017qzi}),
respectively, assuming a cut-and-count analysis.  The current sensitivity of LUX~\cite{Akerib:2016vxi}
is only slightly less than that of XENON1T.
Note that in several of these cases the DM couples differently to protons and neutrons, and is an example
of Isospin-Violating Dark Matter (IVDM)~\cite{Kurylov:2003ra,Giuliani:2005my,Chang:2010yk,Kang:2010mh,Feng:2011vu,Feng:2013fyw}.

\begin{figure}[ht]
\includegraphics*[width=0.49 \textwidth]{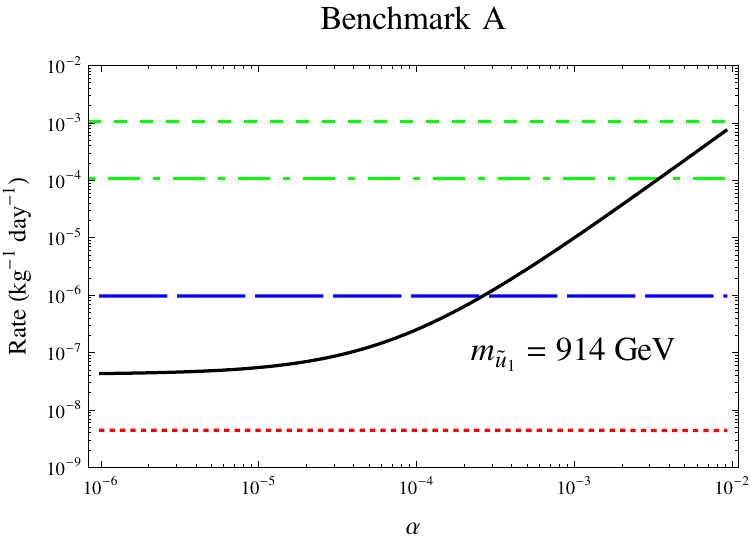}
\includegraphics*[width=0.49 \textwidth]{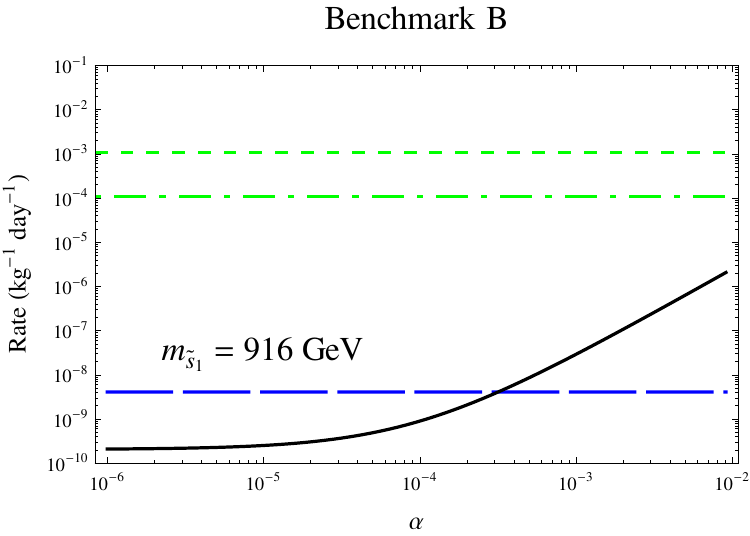} \\ \vspace{2mm}
\includegraphics*[width=0.49 \textwidth]{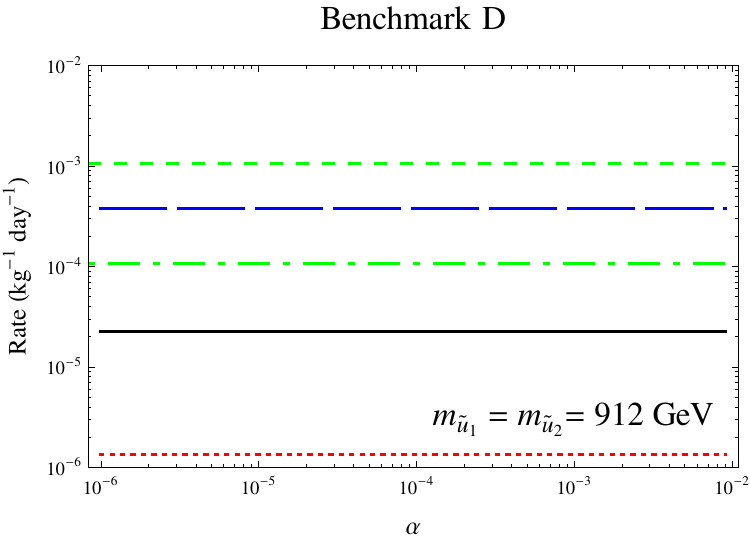}
\includegraphics*[width=0.49 \textwidth]{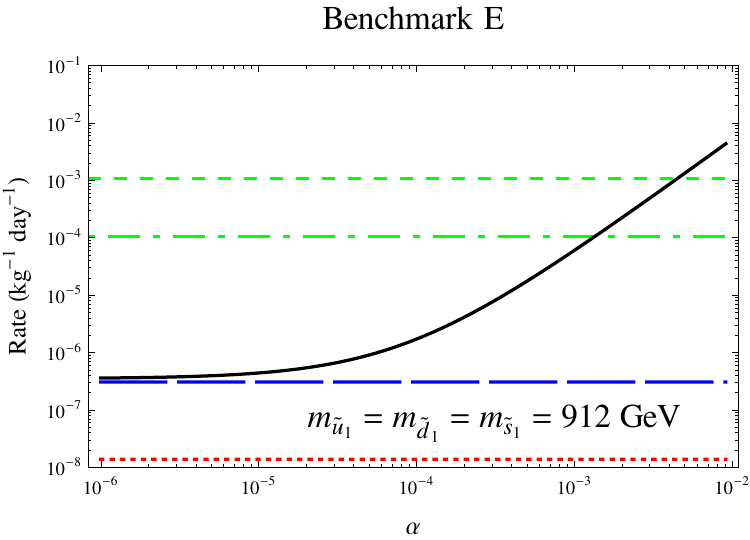}
\caption{
Event rate at a fluorine-based detector (assuming the recoil energy range of PICO-60)
as a function of $\alpha$ for operators ${\cal O}_1$ (dotted red), ${\cal O}_2$ (long-dashed blue) and ${\cal O}_{3,7}$ (solid black)
for a fixed choice of $m_\chi = 900 \gev $.  Shown are Benchmark A (upper left panel),
Benchmark B (upper right panel), Benchmark D (lower left panel)
and Benchmark E (lower right panel).  All other squarks have a negligible effect.
The short-dashed and dash-dotted green lines indicate the current 90\% CL sensitivity PICO-60~\cite{Amole:2017dex} and estimated 90\% CL sensitivity of
PICO-250, respectively.  
For Benchmarks A, B and E,  operator ${\cal O}_7$ begins to
dominate the SI scattering rate for $\alpha \lesssim 10^{-4-5}$, where the
solid black line is flat.
}
\label{fig:ratecompF}
\end{figure}

As expected for all cases except Benchmark D, for large enough $\alpha$, the event rate is dominated by ${\cal O}_3$, which generates
velocity-independent SI scattering (solid black).  Indeed, for large enough $\alpha$, the event rate from ${\cal O}_3$ alone will saturate the upper limit on
the event rate from XENON1T.
The event rate due to ${\cal O}_3$ decreases with $\alpha$, and becomes subleading to that due to ${\cal O}_7$
for $\alpha \lesssim 10^{-4-5}$.
But for Benchmark D, operator ${\cal O}_3$ does not mediate any dark matter scattering, because the degeneracy of the
up-squark masses implies that the necessary  squark mixing can be rotated away.  Instead, for this case, velocity-independent
SI scattering is mediated entirely by ${\cal O}_7$.  The scattering event rates due to operators
${\cal O}_{1,2,7}$ are all largely independent of $\alpha$, since those operators do not flip chirality.
Note that the quark masses
are necessarily a source of chirality mixing, so for $m_\chi \sim {\cal O}(1000~\gev)$, we expect chirality
mixing on the order of at least $10^{-6}$ if dark matter couples to $u$, $d$, and of order
at least $10^{-4}$ if dark matter couples to $s$.

For Benchmarks A, D and E, the rate of scattering in xenon due to ${\cal O}_1$ dominates SD scattering due to ${\cal O}_2$
because there are terms in the matrix element for scattering via ${\cal O}_1$ that receive a coherent enhancement in a
target with a large number of nucleons.
However, scattering due to ${\cal O}_1$ is negligible for Benchmark B because
in that case only a strange-squark is light, and the $s$-quark vector current vanishes for a nucleon state.
But in all cases, velocity-independent SI scattering dominates the event rate even in the $\alpha \rightarrow 0$ limit,
because of the coherent enhancement to scattering via ${\cal O}_7$, which does not flip chirality.

In Figure~\ref{fig:ratecompF}, we similarly plot the event rate in a fluorine-based detector (in the energy $>3.3\kev$ used
by PICO-60) for the four operators, as a function
of $\alpha$, for $m_\chi =900 \gev$.  We consider Benchmarks A, B, D and E, and for each benchmark we adopt the same choice for the
squark masses as in Fig.~\ref{fig:ratecomp}.  The
current 90\% CL sensitivity of PICO-60 ($10^{-3}~\kg^{-1} \da^{-1}$~\cite{Amole:2017dex}) and estimated future
90\% CL sensitivity of
PICO-250 (assumed to be $10^{-4}~\kg^{-1} \da^{-1}$) are plotted as short-dashed  and dash-dotted green lines, respectively.
The qualitative dependence of the rates on $\alpha$ is
the same as for the case of a xenon target, however the rates for scattering mediated by operators ${\cal O}_{1}$  and ${\cal O}_{3,7}$
(dotted red and solid black, respectively) are
smaller, relative to SD scattering (mediated by operator ${\cal O}_2$, long-dashed blue), because the target has fewer nucleons, and thus a smaller coherent-scattering
enhancement.  Indeed, for fluorine there are now regions of parameter space where SD scattering dominates the event rate.
For all cases presented in Fig.~\ref{fig:ratecompF} except Benchmark D, it is again operator ${\cal O}_{3}$ (solid black) that saturates
the upper limit on the event rate from PICO-60 at sufficiently large $\alpha$.

It is interesting to note that the rates of scattering events due to operators ${\cal O}_1$, ${\cal O}_2$ and ${\cal O}_7$ are approximately two
orders of magnitude larger for Benchmark D than for Benchmark A, given our choices of dark matter and squark masses.
The reason is
that for all three of those operators there are terms in the scattering cross section that scale as $Y_R^4$ when a right-handed
squark is light.  Since $Y_R / Y_L = 4$ for an up-type squark, this provides a large enhancement to the scattering rate for Benchmark D.
Indeed, Benchmark D is therefore ruled out by data from XENON1T if $m_{\chi, \tilde u_1, \tilde u_2} \sim 900\gev$.
A similar enhancement could also be expected for Benchmark A, if we had taken $\alpha \sim \pi/2$, in which case there is negligible
squark mixing, but $\tilde q_1 \sim \tilde q_R$.
We will consider the sensitivity of direct detection experiments to dark matter with a relatively
low mass and small mass splitting and will correlate these results with the
thermal relic density in more generality in
the following subsection.

\begin{figure}
\includegraphics*[width=0.49\textwidth]{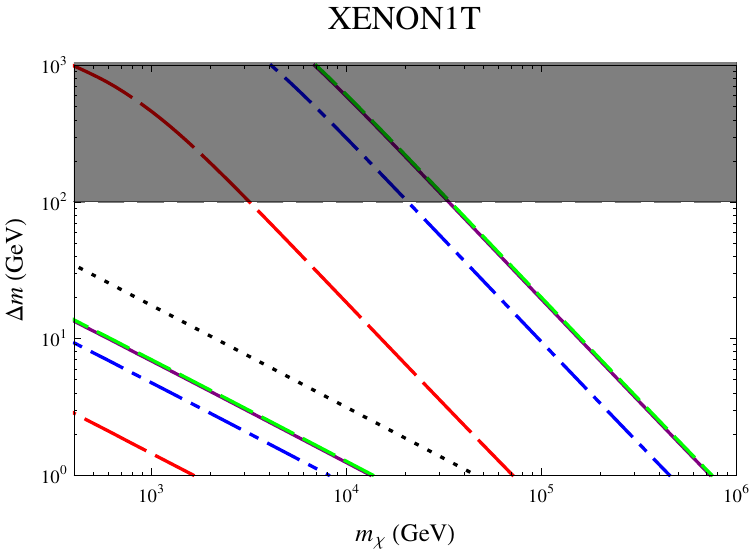}
\includegraphics*[width=0.49\textwidth]{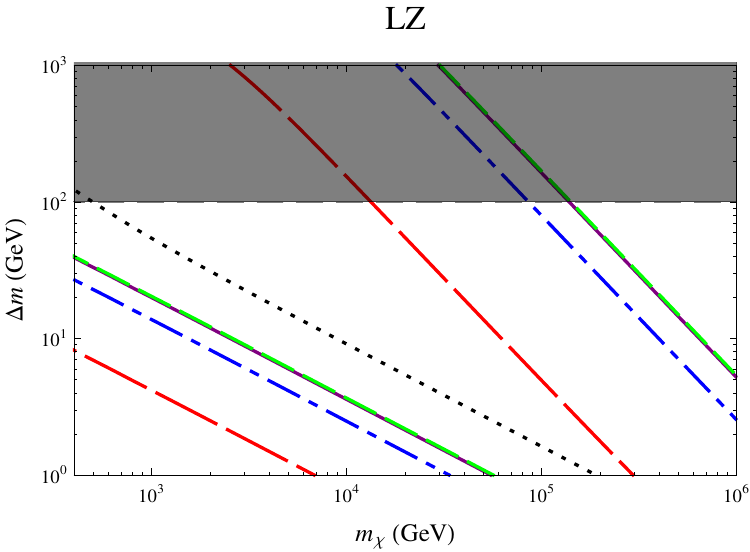} \\ \vspace{2mm}
\includegraphics*[width=0.49\textwidth]{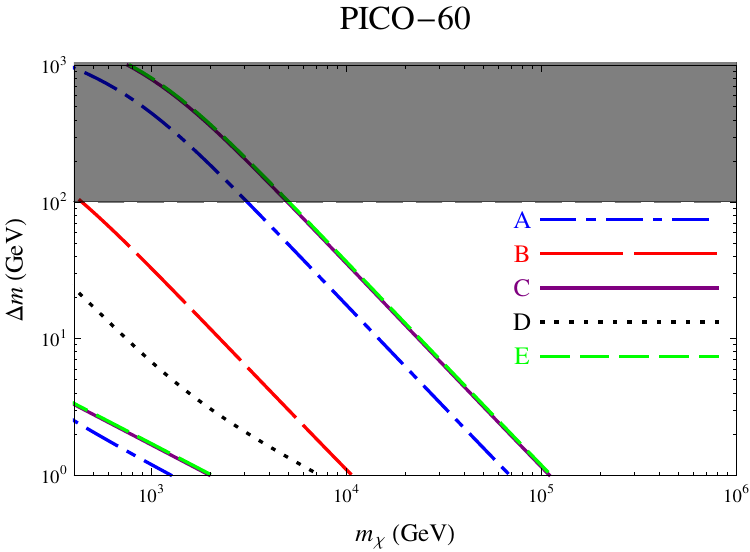}
\includegraphics*[width=0.49\textwidth]{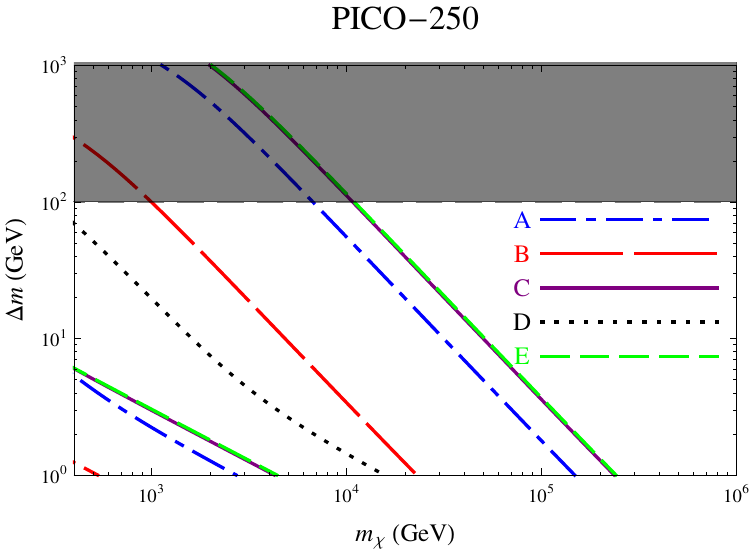}
\caption{Current exclusion and prospective sensitivity (90\% CL) contours in the $(m_\chi, \Delta m)$-plane.  Current
exclusion contours are shown for XENON1T (upper left) and PICO-60 (lower left),
and prospective sensitivity contours are shown for LZ (upper right) and PICO-250 (lower right).
Benchmarks A (dot-dashed blue), B (long-dashed red), C (solid purple), D (dotted black) and E (short-dashed green) are shown.
For Benchmarks A, B, C and E, the upper line is the contour
if $\alpha = \pi/4$, while the lower line is the contour if $\alpha =0$
(for PICO-60, the $\alpha = 0$ contour of Benchmark B is not visible in this region of
parameter space).
For Benchmark D, there is only
one contour because the sensitivity is independent of $\alpha$.  For the grey 
shaded region ($\Delta m > 100\gev$), this analysis is not reliable, as the contributions from 
the heavier squarks cannot be neglected.
}
\label{fig:mchi-m1-plane}
\end{figure}

Figure~\ref{fig:mchi-m1-plane} shows current 90\% CL exclusion contours  for XENON1T and PICO-60,
as well as future 90\% CL sensitivity contours for LZ and PICO-250, in the $(m_\chi, \Delta m)$-plane,
for $m_\chi > 400 \gev$.
We consider Benchmarks
A (dot-dashed blue), B (long-dashed red), C (solid purple), D (dotted black) and E (short-dashed green), for $\alpha =0, \pi/4$.  As expected, the contour for Benchmark C is very similar to that for Benchmark E.
For each contour, the parameter space below the contour is excluded.
For Benchmarks A, B, C and E, the upper contour is for $\alpha = \pi/4$ and
the lower contour is for $\alpha =0$  (for values of $\alpha$ lying between these bounds, the exclusion
contour would lie in between).
For Benchmark D, there is only
one exclusion contour in each panel, because DM scattering is independent of $\alpha$.
For PICO-60, the Benchmark B contour corresponding to $\alpha=0$ does not appear in the plotted parameter space.
Note that, for any choice of $(m_\chi, \Delta m, \alpha)$, all four contact operators can contribute to dark
matter scattering, and the exclusion
contours are determined from the experimental bound on the total scattering event rate, as illustrated in Figures~\ref{fig:ratecomp} and~\ref{fig:ratecompF}.
As we will show in the next subsection, thermal dark matter can only be consistent with the observed relic abundance if $\Delta m \lsim 25 \gev$ and $m_\chi \lsim 1.5 \tev$. Thus, almost all of the allowed parameter space in Figure~\ref{fig:mchi-m1-plane} requires non-thermal production of the observed relic density.

We note that precision electroweak constraints on the $\rho$ parameter can become important at large mixing angles.
However, as discussed at the beginning of Section III, the focus of this work is on scenarios where the light flavor squark mass is quasi-degenerate with the dark matter mass. For such spectra, the $SU(2)_L$ partners of the light flavor squarks we have taken into account for Benchmarks A, B, D and E can safely be decoupled from our direct detection calculations while satisfying constraints on the $\rho$ parameter.
For large mixing angle and $\Delta m \gtrsim 100\gev$, however, the contributions from exchange of $\tilde q_1$ and $\tilde q_2$ are both important.   As we have
seen with Benchmark D, these contributions destructively interfere, leading to a significantly weakened sensitivity.  This analysis is not reliable
in this regime; instead one would need the full spectrum of all squarks.  This region is thus shaded in grey.

Even in the limit $\alpha \rightarrow 0$, XENON1T and PICO-60 already exclude some models for which $m_\chi$ is as large as a few TeV in the degenerate regime, due
to the contribution to the scattering event rate arising from the twist-2 operator.  Not only is this operator not suppressed in the $\alpha \rightarrow 0$ limit (since
it does not mix chiralities), but the energy scale suppression ($m_N^2 / \Delta m^2$) also becomes less severe in the degenerate regime.
The exception is the case where the only light squark is $\tilde{s}_1$ (Benchmark B), in which case experiments are far less sensitive.
As expected, in the degenerate limit XENON1T outperforms PICO-60 at large mixing, and even at small mixing (because of the effect of the twist-2 operator).
The sensitivity of LZ, in the $\alpha \rightarrow 0$ limit, can extend as far
as $\sim 10^5\gev$ (for Benchmark D), with PICO-250 covering much of the same parameter space.
For larger mixing angles, the sensitivities extend much further, especially for LZ.

In the quasi-degenerate limit, the scattering cross section scales as $\mu_A^2 / (m_\chi \Delta m)^2$ and
the scattering rate scales with $m_\chi$ as $m_\chi^{-3}$.  If $\Delta m \sim {\cal O}(m_\chi)$, then the scattering rate instead
scales as $m_\chi^{-5}$.  Thus, an improvement in direct detection
experimental sensitivity produces a greater improvement in mass reach for the quasi-degenerate limit
than for the non-degenerate regime.

That said, it is worth noting that for $m_\chi > 10^3 \gev$ we have $\Delta m / m_\chi < 10^{-3}$ in the degenerate limit.
We have not proposed any mechanism for generating this
level of fine-tuning, so there is no reason to believe that models with $\Delta m$ as small as
$1 \gev$ are natural.
  Nevertheless, this analysis is useful, even in the limit of very small
$\Delta m$ and large $m_\chi$, in determining the level of sensitivity that is possible.

It is also worth noting that although we have considered the sensitivity of direct detection experiments to
models in which the dark matter is a bino and the scalar mediators are squarks, this analysis can
be generalized to other scenarios.  The scenario we've considered corresponds to the choice
$\lambda_{Lq,Rq} = \sqrt{2} g' Y_{L,R}$.  One can rescale the sensitivities given above to any other
scenario by noting that, at maximal mixing, the DM-nucleus scattering cross section is proportional to
$\lambda_L^2 \lambda_R^2$, while for $\alpha =0$ it is proportional to $\lambda_L^4$.

One can also consider the well-studied possibility of searches for
dark matter capture and annihilation to neutrinos in the Sun~\cite{Silk:1985ax,Press:1985ug,Krauss:1985ks}.
However, such searches are only effective if the dark matter annihilation cross section is large in the present
epoch, and if the final state of the annihilation process produces a large number of energetic secondary neutrinos.
But this will not be the case in the scenario we consider here.  In the present epoch, the only relevant annihilation
process is $\chi \chi \rightarrow \bar q q$.  If $q$ is a light quark, then hadronization of the final state will
produce a number of light hadrons which stop in the Sun before decaying, yielding a soft neutrino spectrum which is
difficult to detect.   Searches for this soft neutrino signal at large exposure neutrino detectors can yield a sensitivity
to SD scattering comparable to that of direct detection experiments~\cite{Rott:2012qb,Bernal:2012qh,Rott:2015nma,Rott:2016mzs},
but only for dark matter with $m_\chi \lesssim 10\gev$, and this region of parameter space is already ruled out for this scenario.

\subsection{Relic Density}
\label{sec:rdresults}

In this subsection, we discuss the thermal relic density of dark matter in our benchmark scenarios.
Recent studies of bino-squark co-annihilation have focused on models with a light third generation right-handed squark and have taken into account the effects of Sommerfeld enhancement (for example, see~\cite{deSimone:2014pda,Ibarra:2015nca}) and of bound state formation~\cite{Mitridate:2017izz,Keung:2017kot}.
Such models have qualitatively different features in direct detection searches and the associated enhancements to the total annihilation cross section can significantly alter relic abundance calculations. For the models we consider, the effects of squarkonium formation on the relic density calculation are negligible since the constituent light-flavor squarks will decay before they can form the associated bound state~\cite{Kats:2009bv}. Non-perturbative Sommerfeld QCD corrections can significantly increase the cross section for squark annihilation processes. For instance, if we only consider the dominant annihilation channel in our model, $\tilde q^* \tilde q \to g g$, the Sommerfeld enhancements to the cross section will yield the observed relic density with $\Delta m$ up to $\sim10 \gev$ larger than when considering the perturbative cross section, or alternatively, with a squark mass $\sim 400 \gev$ heavier when $\Delta m =0$~\cite{deSimone:2014pda}.
But while we note that the inclusion of Sommerfeld enhancement would shift the precise scale of the relic density results in $(m_\chi, \Delta m)$ parameter space,
these effects do not change the general features of our analysis.
Detailed implementation of Sommerfeld enhancement in the relic density calculation for a model with
light-flavored squarks is beyond the scope of this work.

\begin{figure}[ht]
\includegraphics*[width=0.49 \textwidth]{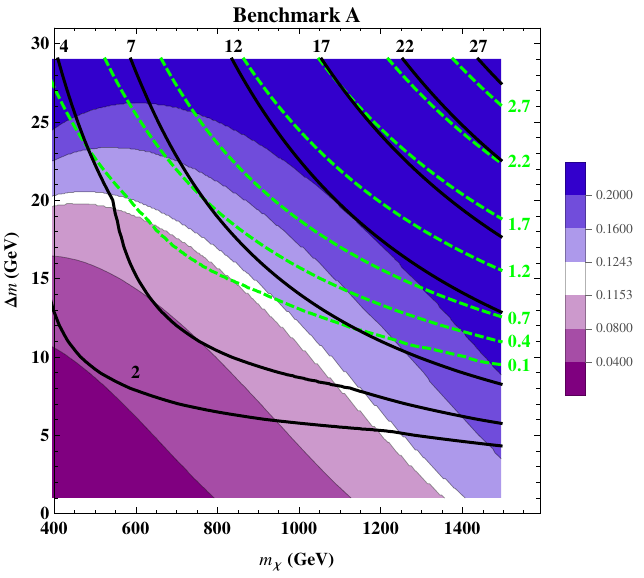}
\includegraphics*[width=0.49 \textwidth]{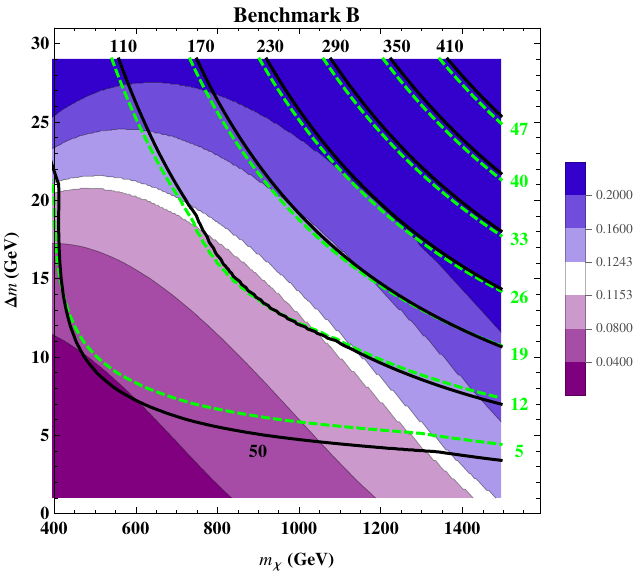} \\ \vspace{2mm}
\includegraphics*[width=0.49 \textwidth]{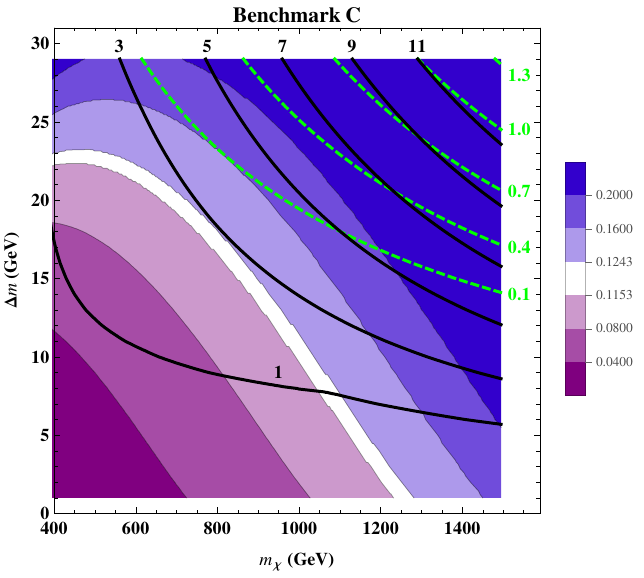}
\includegraphics*[width=0.49 \textwidth]{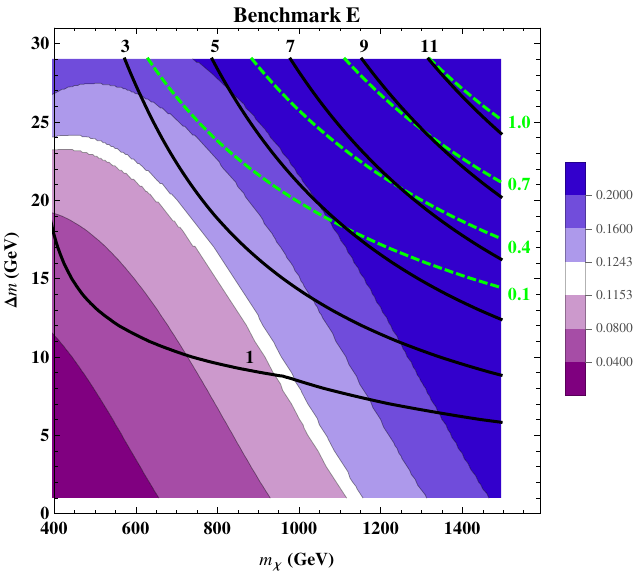}
\caption{Contours of $\Omega h^2$ for Benchmarks A (upper left panel), B (upper right panel),
C (lower left panel), and E (lower right panel),
assuming no L-R mixing ($\alpha = 0$).
The white band corresponds to the region of
parameter space for which the thermal relic density matches the observed DM density.
Also shown are contours of $\alpha$ corresponding to the $90\%$ CL exclusion limit
of XENON1T (black solid) and the expected sensitivity of LZ (green dashed), in units
of $10^{-4}$.  The regions below these contours are excluded.
}
\label{fig:0ud}
\end{figure}

In Figure~\ref{fig:0ud} we show relic density contours for Benchmarks A, B, C and E
in the $(m_{\chi},\Delta m)$ plane, assuming $\alpha = 0$ (no L-R mixing).
We also present contours of $\alpha$ in units of $10^{-4}$ corresponding to the 90\% CL exclusion limits
from XENON1T (black solid) and expected sensitivity of LZ (green dashed).
The excluded regions lie below these contours.  We do not plot contours for either PICO-60 or
PICO-250, as they are in all cases subleading to XENON1T and LZ, respectively.
Note that for $\alpha$ in the range of the plotted contours, the thermal relic density is
indistinguishable from the $\alpha=0$ case.
For regions of parameter space for which
the thermal relic density exceeds the observed dark matter density, we assume that the bino
is produced non-thermally, with a relic density which is equal to the observed dark matter density.
For regions of parameter space for which
the thermal relic density is less than the observed dark matter density,
we instead assume that the bino density is the thermal relic density, with the remainder of the dark matter density arising from some other source.  Direct detection experiments thus have reduced
sensitivity in the region of parameter space where the bino is underabundant, as shown in
Figure~\ref{fig:0ud}.  Had we instead assumed that the bino abundance constituted the entire dark
matter abundance throughout the parameter space, all of the sensitivity contours would
have had the same rough shape.
This shape is controlled by the parametric dependence of the
scattering rate, which for this region of parameter space is $\propto \sigma / m_\chi$.  This rate scales
as $ \propto \alpha^2 / (m_\chi^3 \Delta m^2) $ in the regime where scattering is dominated by operator
${\cal O}_3$, but scales as $ \propto 1 / (m_\chi^3 \Delta m^4)$ at very small $\alpha$, when scattering
is dominated by operator ${\cal O}_7$.

As discussed in Section~\ref{sec:ddresults}, any models with large mixing for which the thermal relic density could match
the observed dark matter density are already ruled out by direct detection experiments.
Note also that the thermal relic density contours for Benchmark B would be the same if the light $s$-squark were replaced by a
$d$-squark, since they have the same charges and couplings.

Since $\alpha$ is small, the $s$-wave term in the
$\chi \chi \to \bar q q$ cross section  vanishes
and the remaining $p$-wave contributions scale as $Y_L^4$.
Thus, only co-annihilation contributions are significant.
Indeed, for $m_{\tilde q_1} \geq 400~\gev$, the ``bulk" region is essentially closed;
if the thermal relic density is to match the observed dark matter density,
then $\Delta m$ must lie in the degenerate region.  As $m_\chi$ increases,
the light scalar must become more and more degenerate with $\chi$ in order to
compensate for the mass-suppression of annihilation/co-annihilation processes,
and the process $\tilde q^* \tilde q \to g g$ dominates the depletion of the relic density even more.
As this is entirely a QCD process, it is flavor-independent.

There is little
difference between Benchmarks A and B for $\alpha =0$, as in this case $\tilde q_1 = \tilde q_L$ and the hypercharge coupling, $Y_L$, is the same for $u$- and $s$-quarks.
But for Benchmark B the correct relic density
is obtained with a slightly larger $m_\chi$ and $\Delta m$ than for Benchmark A.  This difference is due to the marginally larger $Z$-boson coupling to $s$-type squarks, thus enhancing the contribution
from $\tilde q_1 \tilde q_1 \to g Z$ in the case where $q=s$. If, alternatively, we were to assume maximal squark mixing, then the contribution from $\tilde q_1 \tilde q_1 \to g Z$ would be diminished and the effective annihilation cross section would be smaller in general. Relatedly, the relic density for Benchmarks A and B will be more similar assuming maximal mixing than when $\alpha =0$.

For Benchmarks C and E, the $\Delta m$ needed for a model with a
given $m_\chi$ to yield the correct relic density can increase or, perhaps counter-intuitively,
decrease, depending on how small $\Delta m / m_\chi$ is.  This occurs because these benchmarks
have multiple light squarks.
In the most degenerate regions of parameter space, the rates for processes such as $\tilde q^* \tilde q \rightarrow
gg, gZ$ are suppressed because the number density of each squark species is diluted.
Thus, for the region of parameter space with very small $\Delta m$, we see that the correct relic
density is only obtained with
a smaller $m_\chi$ than would be needed in the case in which only $\tilde u_1$ or $\tilde s_1$
(or, equivalently, only $\tilde d_1$) were light.  This effect is more pronounced for Benchmark E,
since there are more light squark species, and the number density of each one is consequently more
heavily diluted.
But at points in parameter space where the bino and squarks are less degenerate, bino-squark co-annihilation
becomes more important.  The dilution of the squark densities causes less of a suppression for co-annihilation at moderate values of $\Delta m / m_\chi$ relative to cases with $\Delta m \to 0$, and the Boltzmann suppression of $\tilde q^* \tilde q \rightarrow gg$ starts to overwhelm squark dilution effects,
leaving the resulting relic depletion rates more similar to sums of those from Benchmarks A and B.

For all of these four benchmarks, though, there is an interesting connection between collider searches, the thermal relic density, and
direct detection.  For each case, there are regions of parameter space in which the dark matter can be a thermal relic and
can escape LHC detection (either because $m_{\chi}$ is beyond the LHC reach or because the $\Delta m$ is too small), but can by
probed by LZ.  But for Benchmarks A and B, if $\alpha$ is sufficiently small
(though it need not be smaller than $m_q / m_\chi$), then there is a region of
parameter space at relatively large $m_\chi$ and moderate $\Delta m$ for which the dark matter can be a thermal relic which evades
detection by both the LHC and LZ.  However, for Benchmarks C and E, LZ can potentially rule out all of the parameter space
in which the dark matter is a thermal relic.

\begin{figure}[ht]
\includegraphics*[width=0.49 \textwidth]{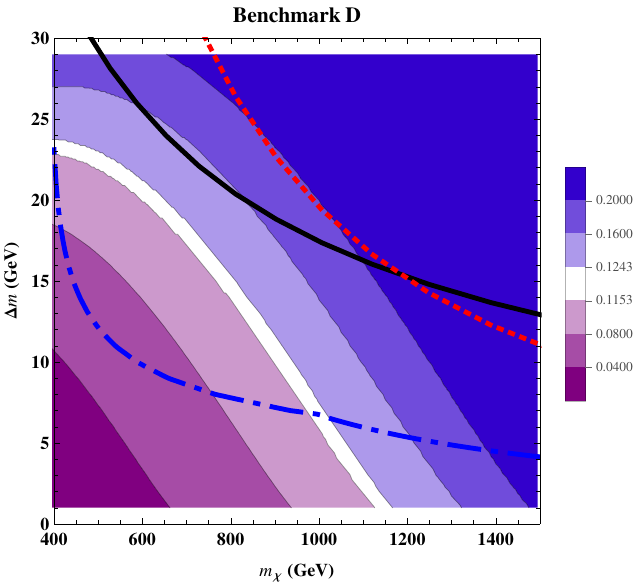}
\caption{Contours of $\Omega h^2$ for Benchmark D.
The white band corresponds to the region of
parameter space for which the thermal relic density matches the observed DM density.
Also plotted are current 90\% CL exclusion
contours from XENON1T (black solid) and PICO-60 (blue dot-dashed), as
well as the expected sensitivity of PICO-250 (red dotted).
}
\label{fig:u12both}
\end{figure}

In Figure~\ref{fig:u12both} we consider Benchmark D ($m_{\tilde u_1} = m_{\tilde u_2}$).
Since the two $u$-type squarks have identical mass, the mixing angle can be rotated away.
For this case, in addition to thermal relic density contours, we plot the current 90\% CL exclusion
contours from XENON1T (black solid) and PICO-60 (blue dot-dashed), as
well as the expected sensitivity PICO-250 (red dotted); the excluded region lies below these contours.
But the entire plotted parameter space lies within the expected sensitivity of LZ.
Again, we assume that, at any point in parameter space, the bino matter density is either the thermal relic density or the
observed dark matter density, whichever is smaller.
For this benchmark we see that the entire region of parameter space in which the
bino thermal relic density can constitute the entire dark matter abundance is already ruled out by XENON1T.

The dilution of the squarks densities also occurs when there are two light $u$-type squarks,
but a difference between this benchmark and the previous cases is that, although the $s$-wave contribution to
$\chi \chi \rightarrow \bar q q$ vanishes for any $\alpha$, the $p$-wave process
receives a large enhancement because a right-handed up-squark is also light and $(Y_R/Y_L)^4 =4^4$. Similarly, bino-squark co-annihilation is enhanced for right-handed squarks relative to left-handed squarks by a factor of $(Y_R/Y_L)^2 =4^2$.
At larger mass differences and lower $m_\chi$, these $p$-wave mixing and co-annihilation contributions are important, and
lead to a relic annihilation rate that is higher than for the case with only a light left-handed
up-squark (Benchmark A). Also, the correct relic density is obtained for slightly lower $m_\chi$ than in Benchmark C because of the previously noted larger coupling of $d$-type squarks to $Z$-bosons, slightly enhancing the rate for $\tilde q_1 \tilde q_1 \to g Z$.

Finally, we note that one can also have unmixed scalars if $\alpha = \pi/2$, in which case the light scalar is $\tilde q_R$.
This scenario is no less well-motivated than $\alpha = 0$, but for reasons of brevity, we simply describe this scenario
qualitatively.
The case with a single $\tilde u_1 = \tilde u_R$ shares qualitative features with Benchmark D. Alternatively, a single right-handed $d$- or $s$-type squark would have smaller enhancements to the mixing and co-annihilation processes relative to Benchmark B since $Y_R/Y_L = 2$, which cannot compensate for the associated suppression of $\tilde q_1 \tilde q_1 \to g Z$ for right-handed squarks.

Note that for all of the benchmark models that we have considered, the dominant
processes which deplete the relic density are independent of $\lambda_{L,R}$.  Although we have focused on the MSSM scenario in which the dark matter
is a bino, the relic density would change very little if we had considered a more general scenario, unless the
$\lambda_{L,R}$ change drastically.  The constraints arising from direct detection experiments depend much more
tightly on the $\lambda_{L,R}$, however, as we have previously discussed.

\section{Conclusion}

We have considered a scenario in which dark matter is a SM gauge-singlet Majorana fermion, coupling
to light SM quarks via exchange of new charged scalar mediators which are nearly degenerate with the dark matter.
Although this simplified model can be
treated as a toy model for an MSSM scenario in which the only light superpartners are a bino-like LSP
and light squarks with a small mass splitting, this scenario has wider applicability.

There are three interesting phenomenological
features in this scenario:
\begin{itemize}
\item{Constraints on the mass of the charged mediators arising from LHC searches can be weakened, due to the
squeezed spectrum.  Although interpretations of LHC squark mass limits directly analogous to
our particular benchmark models are beyond the scope of this work, we assume dark matter and mediators as light as $\sim 400\gev$ may evade current LHC constraints.}
\item{Co-annihilation processes in the early Universe can enhance the relic annihilation rate, increasing
the region of parameter space in which the dark matter could be a thermal relic.  Dark matter as heavy as
$\sim 1500\gev$ can be a consistent thermal relic in this scenario, without accounting for the effects of Sommerfeld enhancement.}
\item{The DM-nucleon scattering cross section (with or without velocity-suppression) is
enhanced by a resonance as the propagator of the mediator goes nearly on-shell.  Future direct detection experiments can probe models in which the dark matter mass is well above 100 TeV.}
\end{itemize}
These features together serve to widen the region of parameter space for which the dark matter candidate can be a
thermal relic, can be probed with direct detection experiments, and can evade current tight constraints from
the LHC.

As more interest has been focused on models with squeezed spectra, new strategies for probing such models
at the LHC have been developed, including the use of spectator jets to give a transverse boost to the
jets and missing transverse momentum.  Some recent work has focused on the use of new kinematic
variables~\cite{Mukhopadhyay:2014dsa},
and on searches for displaced vertices~\cite{Nagata:2017gci}.
It would be interesting to see if further refinements of these
strategies could be used to probe this region of parameter space in which $\chi$ is a good thermal
dark matter candidate.

But it is interesting to note that the sensitivity of future direct
detection experiments can far exceed the maximum
reach of the LHC.  Although this has been known in the limit of maximal scalar mixing, when velocity-independent
SI scattering is dominant~\cite{Kelso:2014qja}, we have found that this is true even for much smaller mixing.
For such models, the rate at which the relic density is depleted is suppressed by the large mass scale of the dark matter,
independent of any co-annihilation.  As a result, such models could only be consistent if the dark matter abundance
was generated non-thermally.  In this limit of heavy dark matter, direct detection experiments, including experiments
like PICO which are focused on SD scattering, could discover not only dark matter, but also the interactions of
QCD-coupled heavy scalars.

The scenario in which direct detection experiments would have the greatest sensitivity is when
the mass splitting $\Delta m / m_\chi$ is less than ${\cal O}(10^{-3})$.  But any such models would appear to
be fine-tuned.  It would be interesting to study if there exist models in which such small mass splittings
occur naturally.

\acknowledgments

We would like to thank Stefania Gori, Wick Haxton, Aaron Pierce and Xerxes Tata for useful discussions,
and CETUP* (Center for Theoretical Underground Physics and Related Areas) for
hospitality and support.
The work of J.~Kumar is supported in part by NSF CAREER Grant No.~PHY-1250573.
The work of P.~Sandick is supported in part by NSF Grant No.~PHY-1417367.
The work of P.~Stengel is supported in part by DOE grant DE-SC007859.
The work of C. Kelso and A. Davidson was supported in part by Space Florida and the National Aeronautics and Space Administration through the University of Central Florida's NASA Florida Space Grant Consortium.

\appendix
\section{Interaction terms for scalar mediator} \label{app:scalInt}

The scalars $\tilde q_{L,R}$ necessarily couple to the $\gamma$, $g$, $Z$ and $W^{\pm}$, as a result of gauge-invariance.
We may write these Lagrangian terms as
\bea
V_{\tilde q} &=& -\imath g_{EM} (\tilde q_{L,R}^* \partial_\mu \tilde  q_{L,R} - \tilde q_{L,R} \partial_\mu \tilde  q^*_{L,R} )
\left( Q  A^\mu - {T_3 - \sin^2 \theta_W Q  \over \sin \theta_W \cos \theta_W} Z^\mu \right)
\nonumber\\
&\,&
 -\imath {g \over \sqrt{2}} (\tilde u_{L}^* \partial_\mu \tilde  d_{L} - \tilde d_{L} \partial_\mu \tilde  u^*_{L} ) W^{+ \mu}  -\imath {g \over \sqrt{2}} (\tilde d_{L}^* \partial_\mu \tilde  u_{L} - \tilde u_{L} \partial_\mu \tilde  d^*_{L} ) W^{- \mu}
\nonumber\\
&\,& + \tilde q_{L,R}^* \left( g_{EM}  Q  A^\mu - g_{EM}  {T_3 - \sin^2 \theta_W Q \over \sin \theta_W \cos \theta_W}  Z^\mu + g_s t^a g^\mu_a \right)
\nonumber\\
&\,& \qquad \qquad \times
\left(  g_{EM} Q  A_\mu -  g_{EM} {T_3 - \sin^2 \theta_W Q \over \sin \theta_W \cos \theta_W}  Z_\mu + g_s t^b g_\mu^b \right)  \tilde q_{L,R}
\nonumber\\
&\,& + {g \over \sqrt{2}} \tilde u_{L}^* \left(  g_{EM} ( Q_u + Q_d )  A_\mu -  g_{EM} \left( {T_{3,u} - \sin^2 \theta_W Q_u \over \sin \theta_W \cos \theta_W} +  {T_{3,d} - \sin^2 \theta_W Q_d \over \sin \theta_W \cos \theta_W} \right)  Z_\mu + g_s t^a g_\mu^a \right) \tilde d_L W^{+ \mu}
\nonumber\\
&\,& + {g \over \sqrt{2}} \tilde d_{L}^*  \left(  g_{EM} ( Q_u + Q_d )  A_\mu -  g_{EM} \left( {T_{3,u} - \sin^2 \theta_W Q_u \over \sin \theta_W \cos \theta_W} +  {T_{3,d} - \sin^2 \theta_W Q_d \over \sin \theta_W \cos \theta_W} \right)  Z_\mu + g_s t^a g_\mu^a \right)  \tilde u_L W^{- \mu}
\nonumber\\
&\,& + { g^2 \over 2}  \tilde q_L^* \tilde q_L W^{\pm}_\mu W^{\mp \mu}
 -\imath g_{s} (\tilde q_{1,2}^* \partial_\mu t^a \tilde  q_{1,2} - \tilde q_{1,2} \partial_\mu t^a \tilde  q^*_{1,2} ) g_a^\mu
\eea
where $Q$ is the squark electric charge and $g_{EM} = g \sin \theta_W$ is the electromagnetic coupling constant.  For $\tilde q_R$, $T_3=0$,
while for $\tilde q_L$ $T_3 = -1/2$ for a down-type squarks and $+1/2$ for an up-type quark.


\begin{thebibliography}{99}

\bibitem{Chang:2013oia}
  S.~Chang, R.~Edezhath, J.~Hutchinson and M.~Luty,
  Phys.\ Rev.\ D {\bf 89}, no. 1, 015011 (2014)
  doi:10.1103/PhysRevD.89.015011
  [arXiv:1307.8120 [hep-ph]].

\bibitem{An:2013xka}
  H.~An, L.~T.~Wang and H.~Zhang,
  Phys.\ Rev.\ D {\bf 89}, no. 11, 115014 (2014)
  doi:10.1103/PhysRevD.89.115014
  [arXiv:1308.0592 [hep-ph]].

\bibitem{Bai:2013iqa}
  Y.~Bai and J.~Berger,
  JHEP {\bf 1311}, 171 (2013)
  doi:10.1007/JHEP11(2013)171
  [arXiv:1308.0612 [hep-ph]].

\bibitem{Papucci:2014iwa}
  M.~Papucci, A.~Vichi and K.~M.~Zurek,
  JHEP {\bf 1411}, 024 (2014)
  doi:10.1007/JHEP11(2014)024
  [arXiv:1402.2285 [hep-ph]].

\bibitem{Garny:2014waa}
  M.~Garny, A.~Ibarra, S.~Rydbeck and S.~Vogl,
  JHEP {\bf 1406}, 169 (2014)
  doi:10.1007/JHEP06(2014)169
  [arXiv:1403.4634 [hep-ph]].

\bibitem{Feng:2008ya}
  J.~L.~Feng and J.~Kumar,
  Phys.\ Rev.\ Lett.\  {\bf 101}, 231301 (2008)
  doi:10.1103/PhysRevLett.101.231301
  [arXiv:0803.4196 [hep-ph]].

\bibitem{Feng:2008mu}
  J.~L.~Feng, H.~Tu and H.~B.~Yu,
  JCAP {\bf 0810}, 043 (2008)
  doi:10.1088/1475-7516/2008/10/043
  [arXiv:0808.2318 [hep-ph]].

\bibitem{Barger:2010ng}
  V.~Barger, J.~Kumar, D.~Marfatia and E.~M.~Sessolo,
  Phys.\ Rev.\ D {\bf 81}, 115010 (2010)
  doi:10.1103/PhysRevD.81.115010
  [arXiv:1004.4573 [hep-ph]].

\bibitem{Fukushima:2011df}
  K.~Fukushima, J.~Kumar and P.~Sandick,
  Phys.\ Rev.\ D {\bf 84}, 014020 (2011)
  doi:10.1103/PhysRevD.84.014020
  [arXiv:1103.5068 [hep-ph]].


\bibitem{Ellis:2012nv}
  J.~Ellis, F.~Luo, K.~A.~Olive and P.~Sandick,
  Eur.\ Phys.\ J.\ C {\bf 73}, no. 4, 2403 (2013)
  doi:10.1140/epjc/s10052-013-2403-0
  [arXiv:1212.4476 [hep-ph]].

\bibitem{Fukushima:2014yia}
  K.~Fukushima, C.~Kelso, J.~Kumar, P.~Sandick and T.~Yamamoto,
  Phys.\ Rev.\ D {\bf 90}, no. 9, 095007 (2014)
  doi:10.1103/PhysRevD.90.095007
  [arXiv:1406.4903 [hep-ph]].


\bibitem{Kelso:2014qja}
  C.~Kelso, J.~Kumar, P.~Sandick and P.~Stengel,
  Phys.\ Rev.\ D {\bf 91}, 055028 (2015)
  doi:10.1103/PhysRevD.91.055028
  [arXiv:1411.2634 [hep-ph]].

  \bibitem{Ellis:1998kh}
  J.~R.~Ellis, T.~Falk and K.~A.~Olive,
  Phys.\ Lett.\ B {\bf 444}, 367 (1998)
  doi:10.1016/S0370-2693(98)01392-6
  [hep-ph/9810360].

\bibitem{Ellis:1999mm}
  J.~R.~Ellis, T.~Falk, K.~A.~Olive and M.~Srednicki,
  Astropart.\ Phys.\  {\bf 13}, 181 (2000)
  Erratum: [Astropart.\ Phys.\  {\bf 15}, 413 (2001)]
  doi:10.1016/S0927-6505(99)00104-8
  [hep-ph/9905481].

\bibitem{Boehm:1999bj}
  C.~Boehm, A.~Djouadi and M.~Drees,
  Phys.\ Rev.\ D {\bf 62}, 035012 (2000)
  doi:10.1103/PhysRevD.62.035012
  [hep-ph/9911496].

\bibitem{Ellis:2001nx}
  J.~R.~Ellis, K.~A.~Olive and Y.~Santoso,
  Astropart.\ Phys.\  {\bf 18}, 395 (2003)
  doi:10.1016/S0927-6505(02)00151-2
  [hep-ph/0112113].


\bibitem{Fukushima:2013efa}
  K.~Fukushima and J.~Kumar,
  Phys.\ Rev.\ D {\bf 88}, no. 5, 056017 (2013)
  doi:10.1103/PhysRevD.88.056017
  [arXiv:1307.7120 [hep-ph]].

\bibitem{Kumar:2016gxq}
  J.~Kumar and C.~Light,
  arXiv:1612.00773 [hep-ph].

\bibitem{Drees:1993bu}
  M.~Drees and M.~Nojiri,
  Phys.\ Rev.\ D {\bf 48}, 3483 (1993)
  doi:10.1103/PhysRevD.48.3483
  [hep-ph/9307208].


\bibitem{Kumar:2013iva}
  J.~Kumar and D.~Marfatia,
  Phys.\ Rev.\ D {\bf 88}, no. 1, 014035 (2013)
  doi:10.1103/PhysRevD.88.014035
  [arXiv:1305.1611 [hep-ph]].

\bibitem{Anand:2013yka}
  N.~Anand, A.~L.~Fitzpatrick and W.~C.~Haxton,
  Phys.\ Rev.\ C {\bf 89}, no. 6, 065501 (2014)
  doi:10.1103/PhysRevC.89.065501
  [arXiv:1308.6288 [hep-ph]].

\bibitem{Hill:2014yxa}
  R.~J.~Hill and M.~P.~Solon,
  Phys.\ Rev.\ D {\bf 91}, 043505 (2015)
  doi:10.1103/PhysRevD.91.043505
  [arXiv:1409.8290 [hep-ph]].

\bibitem{Dienes:2013xya}
  K.~R.~Dienes, J.~Kumar, B.~Thomas and D.~Yaylali,
  Phys.\ Rev.\ D {\bf 90}, no. 1, 015012 (2014)
  doi:10.1103/PhysRevD.90.015012
  [arXiv:1312.7772 [hep-ph]].

\bibitem{Sirunyan:2017cwe}
  A.~M.~Sirunyan {\it et al.} [CMS Collaboration],
  Phys.\ Rev.\ D {\bf 96}, no. 3, 032003 (2017)
  doi:10.1103/PhysRevD.96.032003
  [arXiv:1704.07781 [hep-ex]].

\bibitem{ATLAS:2017dnw}
  The ATLAS collaboration [ATLAS Collaboration],
  ATLAS-CONF-2017-060.


\bibitem{Sirunyan:2017kiw}
  A.~M.~Sirunyan {\it et al.} [CMS Collaboration],
  arXiv:1707.07274 [hep-ex].

\bibitem{Mahbubani:2012qq}
  R.~Mahbubani, M.~Papucci, G.~Perez, J.~T.~Ruderman and A.~Weiler,
  Phys.\ Rev.\ Lett.\  {\bf 110}, no. 15, 151804 (2013)
  [arXiv:1212.3328 [hep-ph]].

\bibitem{deSimone:2014pda}
  A.~De Simone, G.~F.~Giudice and A.~Strumia,
  JHEP {\bf 1406}, 081 (2014)
  doi:10.1007/JHEP06(2014)081
  [arXiv:1402.6287 [hep-ph]].

\bibitem{Bringmann:2007nk}
  T.~Bringmann, L.~Bergstrom and J.~Edsjo,
  JHEP {\bf 0801}, 049 (2008)
  doi:10.1088/1126-6708/2008/01/049
  [arXiv:0710.3169 [hep-ph]].

\bibitem{Bergstrom:1997fh}
  L.~Bergstrom and P.~Ullio,
  Nucl.\ Phys.\ B {\bf 504}, 27 (1997)
  doi:10.1016/S0550-3213(97)00530-0
  [hep-ph/9706232].

\bibitem{Bern:1997ng}
  Z.~Bern, P.~Gondolo and M.~Perelstein,
  Phys.\ Lett.\ B {\bf 411}, 86 (1997)
  doi:10.1016/S0370-2693(97)00990-8
  [hep-ph/9706538].

\bibitem{Ullio:1997ke}
  P.~Ullio and L.~Bergstrom,
  Phys.\ Rev.\ D {\bf 57}, 1962 (1998)
  doi:10.1103/PhysRevD.57.1962
  [hep-ph/9707333].

\bibitem{Kumar:2016cum}
  J.~Kumar, P.~Sandick, F.~Teng and T.~Yamamoto,
  Phys.\ Rev.\ D {\bf 94}, no. 1, 015022 (2016)
  doi:10.1103/PhysRevD.94.015022
  [arXiv:1605.03224 [hep-ph]].

  \bibitem{Belanger:2001fz}
  G.~Belanger, F.~Boudjema, A.~Pukhov and A.~Semenov,
  Comput.\ Phys.\ Commun.\  {\bf 149}, 103 (2002)
  doi:10.1016/S0010-4655(02)00596-9
  [hep-ph/0112278].

  \bibitem{Belanger:2004yn}
  G.~Belanger, F.~Boudjema, A.~Pukhov and A.~Semenov,
  Comput.\ Phys.\ Commun.\  {\bf 174}, 577 (2006)
  doi:10.1016/j.cpc.2005.12.005
  [hep-ph/0405253].

\bibitem{Belanger:2014vza}
  G.~Belanger, F.~Boudjema, A.~Pukhov and A.~Semenov,
  Comput.\ Phys.\ Commun.\  {\bf 192}, 322 (2015)
  doi:10.1016/j.cpc.2015.03.003
  [arXiv:1407.6129 [hep-ph]].

\bibitem{Drees:1990dx}
  M.~Drees and K.~Hagiwara,
  Phys.\ Rev.\ D {\bf 42}, 1709 (1990).
  doi:10.1103/PhysRevD.42.1709


\bibitem{Aprile:2017iyp}
  E.~Aprile {\it et al.},
  arXiv:1705.06655 [astro-ph.CO].

\bibitem{Mount:2017qzi}
  B.~J.~Mount {\it et al.},
  arXiv:1703.09144 [physics.ins-det].


\bibitem{Akerib:2016vxi}
  D.~S.~Akerib {\it et al.} [LUX Collaboration],
  Phys.\ Rev.\ Lett.\  {\bf 118}, no. 2, 021303 (2017)
  doi:10.1103/PhysRevLett.118.021303
  [arXiv:1608.07648 [astro-ph.CO]].


\bibitem{Kurylov:2003ra}
  A.~Kurylov and M.~Kamionkowski,
  Phys.\ Rev.\ D {\bf 69}, 063503 (2004)
  doi:10.1103/PhysRevD.69.063503
  [hep-ph/0307185].

\bibitem{Giuliani:2005my}
  F.~Giuliani,
  Phys.\ Rev.\ Lett.\  {\bf 95}, 101301 (2005)
  doi:10.1103/PhysRevLett.95.101301
  [hep-ph/0504157].


\bibitem{Chang:2010yk}
  S.~Chang, J.~Liu, A.~Pierce, N.~Weiner and I.~Yavin,
  JCAP {\bf 1008}, 018 (2010)
  doi:10.1088/1475-7516/2010/08/018
  [arXiv:1004.0697 [hep-ph]].

\bibitem{Kang:2010mh}
  Z.~Kang, T.~Li, T.~Liu, C.~Tong and J.~M.~Yang,
  JCAP {\bf 1101}, 028 (2011)
  doi:10.1088/1475-7516/2011/01/028
  [arXiv:1008.5243 [hep-ph]].


\bibitem{Feng:2011vu}
  J.~L.~Feng, J.~Kumar, D.~Marfatia and D.~Sanford,
  Phys.\ Lett.\ B {\bf 703}, 124 (2011)
  doi:10.1016/j.physletb.2011.07.083
  [arXiv:1102.4331 [hep-ph]].

\bibitem{Feng:2013fyw}
  J.~L.~Feng, J.~Kumar and D.~Sanford,
  Phys.\ Rev.\ D {\bf 88}, no. 1, 015021 (2013)
  doi:10.1103/PhysRevD.88.015021
  [arXiv:1306.2315 [hep-ph]].

\bibitem{Amole:2017dex}
  C.~Amole {\it et al.} [PICO Collaboration],
  arXiv:1702.07666 [astro-ph.CO].

\bibitem{Silk:1985ax}
  J.~Silk, K.~A.~Olive and M.~Srednicki,
  Phys.\ Rev.\ Lett.\  {\bf 55}, 257 (1985).

  \bibitem{Press:1985ug}
  W.~H.~Press and D.~N.~Spergel,
  Astrophys.\ J.\  {\bf 296}, 679 (1985).

  \bibitem{Krauss:1985ks}
  L.~M.~Krauss, K.~Freese, W.~Press and D.~Spergel,
  Astrophys.\ J.\  {\bf 299}, 1001 (1985).



\bibitem{Rott:2012qb}
  C.~Rott, J.~Siegal-Gaskins and J.~F.~Beacom,
  Phys.\ Rev.\ D {\bf 88}, 055005 (2013)
  doi:10.1103/PhysRevD.88.055005
  [arXiv:1208.0827 [astro-ph.HE]].

\bibitem{Bernal:2012qh}
  N.~Bernal, J.~Martín-Albo and S.~Palomares-Ruiz,
  JCAP {\bf 1308}, 011 (2013)
  doi:10.1088/1475-7516/2013/08/011
  [arXiv:1208.0834 [hep-ph]].


\bibitem{Rott:2015nma}
  C.~Rott, S.~In, J.~Kumar and D.~Yaylali,
  JCAP {\bf 1511}, no. 11, 039 (2015)
  doi:10.1088/1475-7516/2015/11/039
  [arXiv:1510.00170 [hep-ph]].

\bibitem{Rott:2016mzs}
  C.~Rott, S.~In, J.~Kumar and D.~Yaylali,
  JCAP {\bf 1701}, no. 01, 016 (2017)
  doi:10.1088/1475-7516/2017/01/016
  [arXiv:1609.04876 [hep-ph]].


\bibitem{Ibarra:2015nca}
  A.~Ibarra, A.~Pierce, N.~R.~Shah and S.~Vogl,
  Phys.\ Rev.\ D {\bf 91}, no. 9, 095018 (2015)
  doi:10.1103/PhysRevD.91.095018
  [arXiv:1501.03164 [hep-ph]].

\bibitem{Mitridate:2017izz}
  A.~Mitridate, M.~Redi, J.~Smirnov and A.~Strumia,
  arXiv:1702.01141 [hep-ph].

\bibitem{Keung:2017kot}
  W.~Y.~Keung, I.~Low and Y.~Zhang,
  arXiv:1703.02977 [hep-ph].

\bibitem{Kats:2009bv}
  Y.~Kats and M.~D.~Schwartz,
  JHEP {\bf 1004}, 016 (2010)
  doi:10.1007/JHEP04(2010)016
  [arXiv:0912.0526 [hep-ph]].

\bibitem{Mukhopadhyay:2014dsa}
  S.~Mukhopadhyay, M.~M.~Nojiri and T.~T.~Yanagida,
  JHEP {\bf 1410}, 12 (2014)
  doi:10.1007/JHEP10(2014)012
  [arXiv:1403.6028 [hep-ph]].

\bibitem{Nagata:2017gci}
  N.~Nagata, H.~Otono and S.~Shirai,
  JHEP {\bf 1703}, 025 (2017)
  doi:10.1007/JHEP03(2017)025
  [arXiv:1701.07664 [hep-ph]].




\end{thebibliography}
\end{document}